\documentclass[journal]{IEEEtran}
\usepackage{cite}
\usepackage{amsmath,amssymb,amsfonts}
\usepackage{algorithmic}
\usepackage{graphicx}
\usepackage{epstopdf} 
\usepackage{textcomp}
\usepackage{multirow}
\usepackage{subfigure}
\usepackage{color,xcolor}
\usepackage{makecell}
\def\BibTeX{{\rm B\kern-.05em{\sc i\kern-.025em b}\kern-.08em
		T\kern-.1667em\lower.7ex\hbox{E}\kern-.125emX}}
\begin{document}
	\title{Wireless Channel Measurements and Characterization in Industrial IoT Scenarios}
	
	\author{Li Zhang, \IEEEmembership{Student Member, IEEE}, Cheng-Xiang Wang, \IEEEmembership{Fellow, IEEE}, Zihao Zhou, \IEEEmembership{Student Member, IEEE}, Yuxiao Li, Jie Huang, \IEEEmembership{Member, IEEE}, Lijian Xin, \IEEEmembership{Member, IEEE}, Chun Pan, Dabo Zheng, and \\ Xiping Wu, \IEEEmembership{Senior Member, IEEE}
		\thanks{Manuscript received 12 October, 2023; revised 8 April, 2024, and 20 July, 2024; accepted 25 September, 2024. This work was supported by the National Key R\&D Program of China under Grant 2021YFB2900300, the National Natural Science Foundation of China (NSFC) under Grants 61960206006 and 62271147, the Fundamental Research Funds for the Central Universities under Grant 2242022k60006, the Key Technologies R\&D Program of Jiangsu (Prospective and Key Technologies for Industry) under Grants BE2022067, BE2022067-1, BE2022067-3, and BE2022067-4, the Start-up Research Fund of Southeast University under Grant RF1028623029, and the Research Fund of National Mobile Communications Research Laboratory, Southeast University, under Grant 2024A05. The review of this article was coordinated by Prof. Vittorio Degli-Esposti. (\emph{Corresponding Authors: Cheng-Xiang Wang and Jie Huang})}
		\thanks{L. Zhang, C.-X. Wang, Z. Zhou, and J. Huang are with the National Mobile Communications Research Laboratory, School of Information Science and Engineering, Southeast University, Nanjing 211189, China, and also with the Purple Mountain Laboratories, Nanjing 211111, China (email: \{li-zhang, chxwang, zhouzihao, j\_huang\}@seu.edu.cn).}
		\thanks{Y. Li is with the Nanjing Jiangning Power Supply Company of State Grid, Nanjing 210000, China (email: liyx13@js.sgcc.com.cn).}		
		\thanks{L. Xin is with the Purple Mountain Laboratories, Nanjing 211111, China (email: xinlijian@pmlabs.com.cn).}
		\thanks{C. Pan is with Huawei Technologies Co., Ltd., Nanjing 210012, China (e-mail: panchun@huawei.com).}
		\thanks{D. Zheng is with NIO Technologies Co., Ltd., Nanjing 210046, China (e-mail: alen.zheng@nio.com).}
		\thanks{X. Wu is with the School of Information Science and Engineering, Southeast University, Nanjing 211189, China, and also as a visiting professor with the School of Electrical and Electronic Engineering, University College Dublin, Dublin, D04 V1W8, Ireland (e-mail: xiping.wu@seu.edu.cn).}
	}
	\IEEEpubid{\begin{minipage}{\textwidth}\ \\[30pt] \centering
			Copyright \copyright 2024 IEEE. Personal use of this material is permitted. 
			However, permission to use this material for any other purposes must \\ be obtained 
			from the IEEE by sending an email to pubs-permissions@ieee.org.
	\end{minipage}}
	
	\markboth{IEEE TRANSACTIONS ON VEHICULAR TECHNOLOGY,~Vol.~xx, No.~xx, MONTH~2024}%
	{Shell \MakeLowercase{\textit{et al.}}: Bare Demo of IEEEtran.cls for IEEE Journals}
	
	\maketitle
	
	\begin{abstract}
	Wireless Fidelity (Wi-Fi) communication technologies hold significant potential for realizing the Industrial Internet of Things (IIoT). In this paper, both Single-Input Single-Output (SISO) and polarized Multiple-Input Multiple-Output (MIMO) channel measurements are conducted in an IIoT scenario at the less congested Wi-Fi band, i.e., 5.5~GHz. The purpose is to investigate wireless characteristics of communications between access points and terminals mounted on automated guided vehicles as well as those surrounding manufacturing areas. For SISO channel measurements, statistical properties including the delay Power Spectral Density (PSD), path loss, shadowing fading, delay spread, excess delay, K-factor, and amplitude distribution of small-scale fading are analyzed and compared with those observed in an office scenario. For MIMO channel measurements, results show that there are multiple Dense Multipath Component (DMC) processes in the delay PSD. An estimation algorithm based on the algorithm for a single DMC process is proposed to effectively process the multi-processes data. Moreover, delay, angular, power, and polarization properties of DMCs are investigated and compared with those of specular multipath components. Furthermore, effects of DMCs on Singular Values (SVs) and channel capacities are explored. Ignoring DMCs can overestimate SVs and underestimate channel capacities.	
	\end{abstract}
	
	\section{Introduction}
	\IEEEPARstart{A}{dvancements} in Internet of Things (IoT) technologies---identification~\cite{Liu2022}, communication~\cite{Nguyen2022,Wang2022}, computation~\cite{Li2022}, sensing, etc.---have propelled Industrial IoT (IIoT) applications. IIoT enables the seamless integration of sensors, controllers, and various devices into industrial production processes, enhancing automation and efficiency. In manufacturing scenarios, Automated Guided Vehicles (AGVs), mobile robots, and robotic arms are indispensable. Hence, wireless networks are preferred for those mobile devices by offering flexible deployment, convenient maintenance, and low costs. Meanwhile, Customer Premises Equipment (CPE) can extend network accessibility, integrating legacy  equipment into wireless networks.
	
	Amidst wireless communication technologies, Wireless Fidelity (Wi-Fi) emerges as a versatile solution for industrial networks due to its utilization of unlicensed spectrum and the extensive existing infrastructure. The latest Wi-Fi 6 is specifically designed for high-density and high-capacity services. It operates seamlessly across 2.4~GHz and 5~GHz bands, leveraging advantages of each band. Notably, the 5~GHz band offers a larger number of non-overlapping channels and can therefore enhance network capacities and reduce interferences. Other newly introduced technologies such as orthogonal frequency-division multiple access, target wake time, etc., can significantly improve the spectral efficiency~\cite{IEEE}. Therefore, strategic deployment of Wi-Fi 6 alongside CPE enables wireless connectivity of industrial devices. However, IIoT scenarios are characterized by higher ceilings, larger floor spaces, and more intricate obstructions~\cite{Jiang2020-2}. Wireless communications between Access Points (APs) and CPE mounted on AGVs or production machinery can encounter intricate multipath effects. To successfully design and deploy Wi-Fi networks in IIoT scenarios, conducting channel measurements and characteristic analysis at Wi-Fi bands is necessary~\cite{Wang2020}.
	
	The existing channel measurements in IIoT scenarios are summarized in \mbox{Table~\ref{tab_IIoT_CMM}}. Basically, the measured frequency bands can be classified as five types, i.e., the Fifth Generation (5G) New Radio (NR) bands, not special bands, Industrial Scientific Medical (ISM) bands, Ultra-WideBand (UWB) bands, and Wi-Fi bands. For 5G NR bands, channel measurements were conducted at 3.5~GHz~\cite{Zhong2020,Xu2019,Adegoke2019}, 3.7~GHz~\cite{Schmieder2020}, 3.75~GHz~\cite{Burmeister2021}, 4.1~GHz~\cite{Kim2021}, 4.9~GHz~\cite{Ying2020,Jiang2020,Guan2020}, and  28~GHz~\cite{Ying2020,Guan2020,Wang2021,Ito2022,Solomitckii2019,Schmieder2019,Schmieder2020}.  The investigated channel characteristics mainly focused on Path Loss (PL), Shadowing Fading (SF), Delay Spread (DS), delay Power Spectral Density (PSD), and Angular Spreads (ASs) at milimeter wave bands. For not special bands, Multiple-Input Multiple-Output (MIMO) channel measurements were conducted at 3~GHz~\cite{Tanghe2014}, 1.35~GHz~\cite{Gaillot2015}, and 1.3~GHz~\cite{Hanssens2018}. Results in~\cite{Tanghe2014} found that the power of Dense Multipath Components (DMCs) accounted for 23\% to 70\% of the total channel power and the reverberation time was nearly constant. In~\cite{Gaillot2015}, polarized properties of Specular Multipath Components (SMCs) and DMCs in a large industrial hall were compared. Authors in~\cite{Hanssens2018} analyzed the delay PSD, Cross-Polarization Ratio (XPR), DMC power ratio, and reverberation time of different polarization pairs. These measurements highlighted a significant effect of DMCs on the total power and investigated the polarized properties of DMCs. However, angular properties of DMCs and their effects on the system performance are not explored. Besides, channel characteristics at these frequency bands cannot accurately describe those at Wi-Fi bands.
	
	In terms of Wi-Fi frequency bands, i.e., 2.4--2.4845~GHz and 5.15--5.85~GHz, several channel measurements at ISM bands~\cite{Croonenbroeck2017,Ai2015,Ai2015-2,Narrainen2019,Tanghe2010} and UWB bands~\cite{Razzaghpour2019, Karedal2004} covered part Wi-Fi bands. The interested ISM bands are 915~MHz, 2.45~GHz, and 5.8~GHz. Authors in~\cite{Croonenbroeck2017} conducted Single-Input Single-Output (SISO) channel measurements at 5.8~GHz in various industrial environments and analyzed results about the PL, DS, and coherence bandwidth. In~\cite{Ai2015}, frequency domain channel measurements were performed from 800~MHz to 2.7~GHz to study the PL, SF, and their effects on channel capacity. In~\cite{Ai2015-2,Narrainen2019,Karedal2004,Tanghe2010}, Virtual Antenna Arrays (VAAs) were employed to investigate the amplitude distribution of the Small-Scale Fading (SSF). The Salen--Valenzula (S--V) model was adopted in the UWB communication with parameters obtained from measurements. Authors in~\cite{Razzaghpour2019} performed SISO measurements at 3--8 GHz and channel properties including PL, SF, DS, and Excess Delay (ED) were analyzed. Authors in~\cite{Jaeckel2019} performed three types of MIMO channel measurements at 2.37~GHz and 5.4~GHz to obtain statistical distributions of the eight Large-Scale Parameters (LSPs), such as K-factor (KF), DS, etc. In~\cite{Traore2018}, MIMO channel measurements were conducted in an aircraft manufacturing plant at 2.4~GHz and 5.4~GHz. Two exponential functions were employed to model the delay PSD. However, antenna heights in these channel measurements are nearly the same, while~\cite{Tanghe2010} and~\cite{Jaeckel2019} imitate communications between AP and terminals. Besides, investigated statistical properties of these channel measurements are incomplete and exclude characteristics of DMCs.
		
	In summary, existing channel measurements in IIoT scenarios mainly focus on characterizing 5G NR  bands, while only a few works are related to Wi-Fi bands, especially the less congested 5~GHz Wi-Fi band (5.15--5.85~GHz). Antenna heights in these channel measurements are nearly the same, which cannot provide an accurate analysis of channel characteristics for the scenario between APs and industrial terminals, particularly in AP-AGV scenario. In the AGV communication situation, the receiving antenna is set at a very low height, leading to more severe multipath fading compared to that at higher antenna heights. In addition, current investigations of channel characteristics at 5 GHz Wi-Fi bands are incomplete, particularly in studying the LSF, angular, polarization, and DMCs characteristics. To fill the above gaps, we conduct both SISO and MIMO channel measurements at 5.5~GHz in an automobile manufactory. Two typical Wi-Fi application scenarios are considered, i.e., communications between APs and industrial CPEs installed on AGVs and manufacturing equipment. Based on channel measurements, a comprehensive characterization is conducted and compared with results in an indoor office scenario~\cite{Zhang2023}. The main contributions and novelties of this paper are summarized as follows.		
	\begin{enumerate}
		
		\item{Both SISO and MIMO channel measurements and characteristics analysis are conducted at 5.5~GHz between APs and terminals in IIoT application scenarios. The channel measurement dataset, along with channel characterization, not only furnishes parameters essential for IIoT channel models but also provides guidance for designing and optimizing Wi-Fi systems in IIoT scenarios.}
		
		\item{A novel estimation algorithm is proposed to accommodate the multiple DMC processes in the delay PSD, which is a unique property in the measured IIoT scenario. The first-order difference method is applied to detect transition points in the slope of each process and base delays are normalized so that they can be iterated under an equivalent number of frequency points. The proposed algorithm enhances the power extraction efficiency from 79\% to 99.6\%, thus exhibiting superior performance over the existing algorithms with a single DMC process.} 
		
		\item{A comprehensive channel characteristic analysis is conducted. For SISO channel measurements, statistical properties including the delay PSD, PL, SF, DS, ED, KF, and amplitude distribution of SSF are studied. Regarding MIMO channel measurements, the delay, angular, power, and polarization characteristics of SMCs and DMCs are analyzed and compared. Furthermore, effects of DMCs on Singular Values (SVs) and channel capacities are explored. Results show that incorporating both SMCs and DMCs leads to a decrease in SVs and an increase in channel capacities compared to results with only SMCs.}
	\end{enumerate}
	
	The rest of this paper is structured as follows. Section~\ref{sec:Channel Measurements} outlines the channel measurement campaign in IIoT scenarios. In Section~\ref{sec:measurement data processing}, the method of estimating channel parameters of SMCs and DMCs is described. Measurement results and analysis are presented in Section~\ref{sec:Results and Analysis}. Section~\ref{sec:conclusions} draws conclusions of this paper.
	\begin{table*}
		\begin{center}	
			\caption{Summary of Channel measurements in IIoT scenarios.}
			\label{tab_IIoT_CMM}
			\setlength{\tabcolsep}{1.5pt}
			\begin{tabular}{|c|c|c|c|c|c|c|}
				\hline
				\textbf{Ref.} & \textbf{Environment} & \makecell[c]{\textbf{Frequency} \\ \textbf{(GHz)}} & \makecell[c]{\textbf{Bandwidth}\\ \textbf{(MHz)}} & \textbf{\makecell[c]{Antenna\\ Configuration}} & \textbf{\makecell[c]{Antenna \\Height (m)}} &\textbf{Channel Properties} \\
				\hline	
				~\cite{Zhong2020}&\makecell[c]{Automobile \\assembly factory} &3.5 &160 &64$\times$64 MIMO &Tx: 4.1, Rx: 1.5 &\makecell[c]{Delay PSD, DS, ED, ASD, ASA,\\ ESD, ESA, channel rank} \\
				\hline				
				~\cite{Xu2019}&Large workshop &3.5 &160 &64$\times$64 MIMO &Tx: 3 &Delay PSD, amplitude of SSF, DS\\
				\hline
				~\cite{Adegoke2019}&Machine workshop &3.4--3.8 &400 &SIMO (VAA) & / &Power decay rate, delay PSD, DS\\
				\hline			
				~\cite{Schmieder2020}&\makecell[c]{Machining\\ workshop hall} &3.7, 28 &2000 &\makecell[c]{3.7: SISO\\ 28: SIMO (VAA)} &Tx: 1.85, Rx: 1.44 &\makecell[c]{PL, SF, DS\\ delay PSD, angular PSD, ASA}\\
				\hline	
				~\cite{Burmeister2021}&AGV to AGV &3.75 &100 &SISO &Tx, Rx: 0.8 &Received power, DS\\
				\hline
				~\cite{Kim2021}&Manufacturing plant &4.1 &100 &SISO &Tx: 3.7, Rx: 1.5 &PL, SF, DS, KF\\
				\hline
				~\cite{Ying2020,Jiang2020,Guan2020}&Factory &\makecell[c]{4.9\\ 28} &\makecell[c]{100\\ 800} &SISO &\makecell[c]{Tx:2.5, Rx: 2.5\\Tx:2.5, Rx: 1.9\\Tx:1.9, Rx: 1.9\\Tx:1.9, Rx: 0.9} & PL, SF, DS, KF\\
				\hline	
				~\cite{Wang2021} & \makecell[c]{Fiber optic \\cable laboratory} & 28 & 2400 & \makecell[c]{SISO \& SIMO\\ (VAA)}& \makecell[c]{Tx, Rx: 0.84 \\Tx, Rx: 1.6} &PL, SF, DS, angular PSD, ASA \\
				\hline
				~\cite{Ito2022}&Japanese sake factory &28 &100 &\makecell[c]{SISO \& SIMO (VAA)} &Tx: 2.5, Rx: 1.8 &\makecell[c]{PL, SF, DS, angular PSD, ASA, ESA}\\	
				\hline
				~\cite{Solomitckii2019}&Manufacturing zone &28 &2000 &SIMO (VAA) &Rx: 1.6 &PL, SF, DS, ASA, XPR\\
				\hline	
				~\cite{Schmieder2019}&Circle-shaped machine hall &28 &2000 &SIMO (VAA) &Tx: 7, Rx: 1.7\&4 &\makecell[c]{PL, SF, DS, delay PSD\\ angular PSD, ASD, ASA}\\	
				\hline	
				~\cite{Tanghe2014}&Workshop for reparation &3 &100 &\makecell[c]{8$\times$8 MIMO (VAA)} &Tx: 1.5, Rx: 1.5 &\makecell[c]{DMC power ratio and reverberation time}	\\
				\hline	
				~\cite{Gaillot2015}&Large industrial hall &1.3 &22 &\makecell[c]{12$\times$12 MIMO (VAA)} &Tx: 1.6, Rx: 1.6 &\makecell[c]{ DS, XPR, DMC power ratio}\\
				\hline
				~\cite{Hanssens2018}&Flower auction warehouse &1.35 &80 &8$\times$8 MIMO &Tx: 6, Rx: 2 &\makecell[c]{DMC reverberation time, \\XPR, delay PSD, DMC power}\\
				\hline					
				~\cite{Croonenbroeck2017}&\makecell[c]{Warehouse,\\ manufacturing shop, \\mechanical workshop} &5.8 &600 &SISO &\makecell[c]{Tx, Rx: 1.9\\Tx: 1.57, Rx: 1.85\\Tx: 1.57, Rx: 1.4} &PL, DS, coherence bandwidth\\
				\hline
				~\cite{Ai2015}&\makecell[c]{Assembly shop, electronics\\ room, mechanical room} &0.8--2.7 &1900 &SISO &Tx: 1.8, Rx: 1.8 &\makecell[c]{PL, SF, channel capacity}\\
				\hline	
				~\cite{Ai2015-2}&\makecell[c]{Assembly shop, electronics\\ room, mechanical room} &0.8--2.7 &1900 &\makecell[c]{5$\times$18 MIMO \\(VAA)} &Tx: 1.8, Rx: 1.8 &\makecell[c]{Delay PSD, power decay rate, \\amplitude of SSF}\\
				\hline							
				~\cite{Narrainen2019}&Machinery room &2-6 &4000 &\makecell[c]{9$\times$27 MIMO\\ (VAA)} &/ &\makecell[c]{Power decay rate}\\
				\hline			
				~\cite{Tanghe2010}&\makecell[c]{Production rooms} &0.8--4 &3200 &\makecell[c]{4$\times$12 MIMO (VAA)}&Tx: 6, Rx: 2 &\makecell[c]{Power decay rate, amplitude of SSF}\\
				\hline
				~\cite{Karedal2004}&Factory hall &3.1--10.6 &7500 &\makecell[c]{7$\times$7 MIMO (VAA)} &Tx: 1, Rx: 1 &\makecell[c]{Power decay rate, amplitude of SSF}\\
				\hline
				~\cite{Razzaghpour2019}&Smart production lab &3--8 &5000 &SISO &Tx, Rx: 1--2 &PL, SF, DS, ED\\
				\hline	
				~\cite{Jaeckel2019}&Five factory halls &\makecell[c]{5.4\\ 2.37} &\makecell[c]{200\\ 50} &\makecell[c]{32/30$\times$50 MIMO\\ 32$\times$56 MIMO} &Tx: 2--8, Rx: 2 &\makecell[c]{SF, DS, KF, ASD,\\ ASA, ESD, ESA, XPR}\\											
				\hline
				~\cite{Traore2018}&\makecell[c]{Aircraft manufacturing plant} &2.4, 5.4 &\makecell[c]{100 \& 500} &\makecell[c]{4$\times$4 MIMO (VAA)} &Tx: 1.5, Rx: 1.5 &Delay PSD, DS\\
				\hline							
				\makecell[c]{This \\paper}&Automobile manufactory &5.5 &320 &
				\makecell[c]{SISO \&\\32$\times$64 MIMO} &\makecell[c]{Tx: 4.8, Rx: 0.3 \& 1\\Tx: 4.5, Rx: 1.35} &\makecell[c]{SISO: delay PSD, PL, SF, DS, ED, KF, \\and amplitude of SSF \\MIMO: delay, angular, power, and\\ polarization properties of SMCs and DMCs, \\effects of DMCs on SVs and capacity}\\
				\hline															
			\end{tabular}
		\end{center}	
	\end{table*}	
	
	\section{Channel Measurements}
	\label{sec:Channel Measurements}	
	\subsection{Measurement Environment}
\begin{figure}[!t]
		\centering
		{\includegraphics[width=0.8\columnwidth]{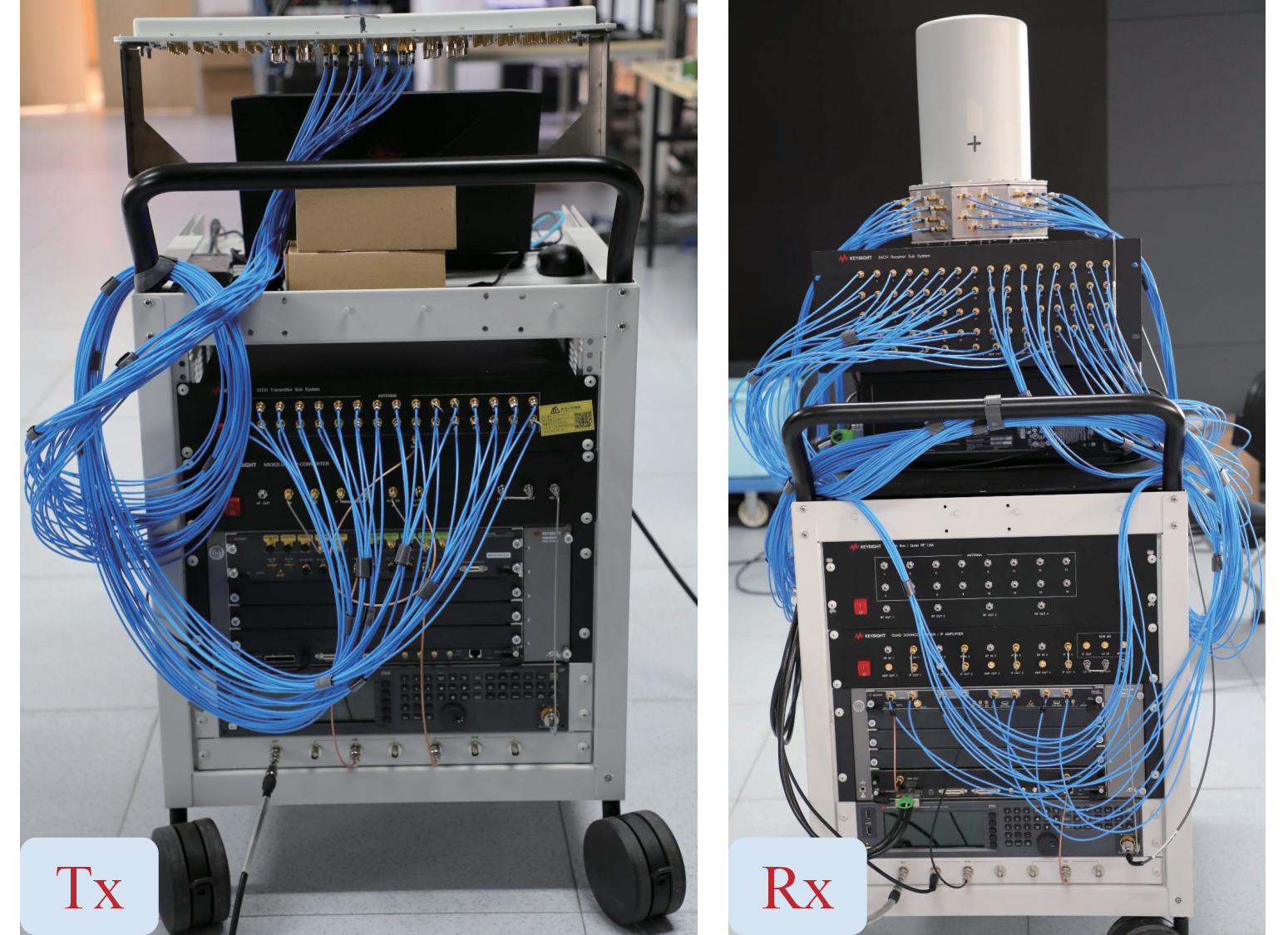}}
		\caption{Time domain 32$\times$64 MIMO channel sounder.}
		\label{equipment}
	\end{figure}

	Channel measurements are performed by a time domain channel sounder, as depicted in Fig.~\ref{equipment}. It includes a Transmitter (Tx) and a Receiver (Rx) and more detailed information of this sounder can be found in~\cite{Zhang2023}. 
		
\begin{figure*}[!t]
		\centering
		\subfigure[]{\includegraphics[width = 1.65\columnwidth]{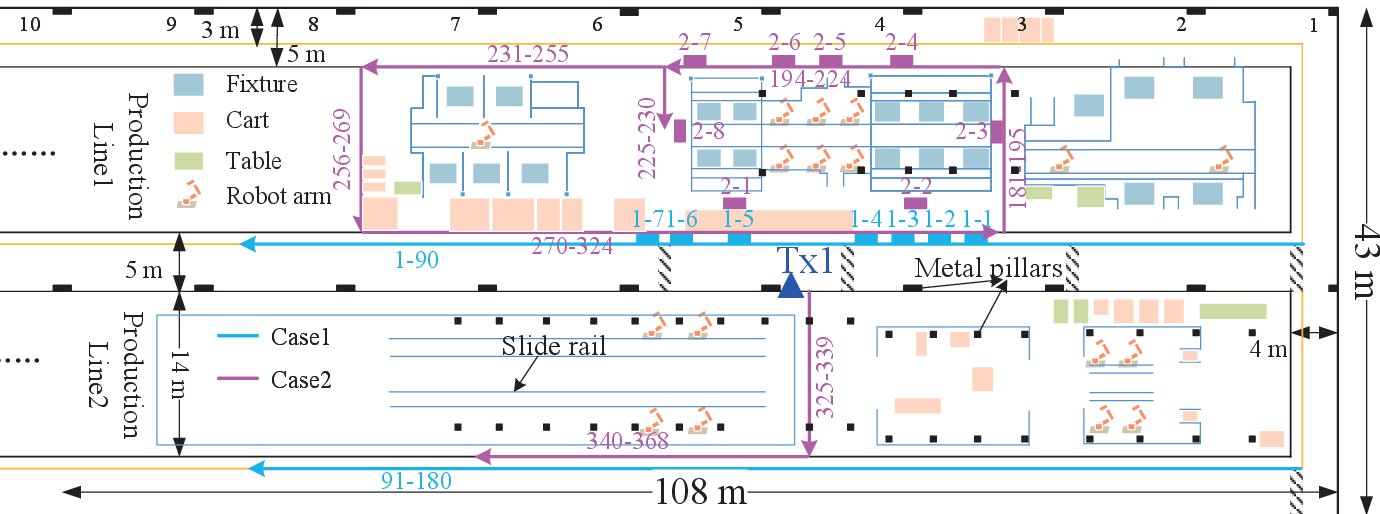}}		
		\subfigure[]{\includegraphics[width = 1.65\columnwidth]{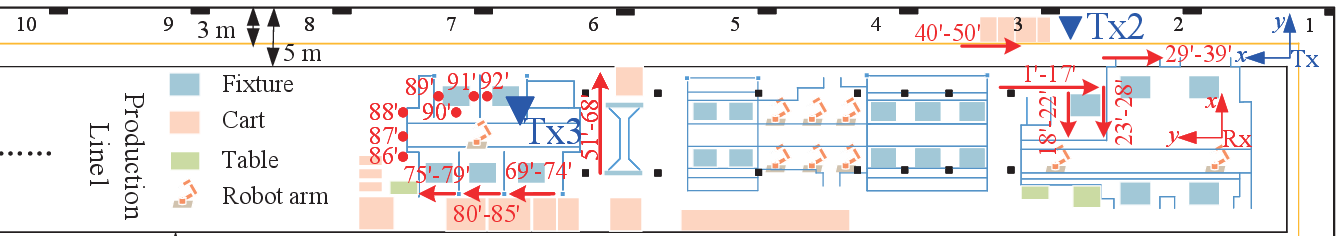}}
		\caption{Measurement positions of (a) SISO and (b) MIMO channel measurements in IIoT scenarios.}
		\label{IIoT_measurement_case}	
	\end{figure*}

	The measurement campaigns are conducted in a large automobile manufactory with a size of \mbox{156$\times$100$\times$14.6~m$^3$}. This manufactory can be primarily divided into an idle section and the other section comprising three assembly lines. Three channel measurement cases are conducted in both Line-Of-Sight (LOS) and Non-LOS (NLOS) scenarios within the area of assembly lines. Due to the restrictions, photography was not permissible during measurements. Fig.~\ref{IIoT_measurement_case} presents a layout of the measurement area and offers a concise depiction. Each assembly line is 156$\times$14~m$^2$ and consists of several operating islands equipped with operating platforms, mechanical arms, slide rails, and metal materials. Between the assembly lines, there are pedestrian walkways with a width of 5~m. On one side of each assembly line, metal pillars are evenly distributed and APs are spaced on these metal pillars. In addition, numerous irregularly shaped steel grids exist on the ceiling and the ground floor is very smooth. 
	
	\subsection{Measurement Setup}
	For SISO channel measurements, two identical omnidirectional antennas are used at Tx and Rx sides to obtain characteristics of the PL, DS, ED, etc. The omnidirectional antenna works at 2--8~GHz with a gain of 0 dBi. The radiation pattern of the omnidirectional antenna exhibits a wide elevation beam with a 3~dB beamwidth of 140$^\circ$. Here, omnidirectional antennas are employed to analyze propagation channel characteristics independent of the antenna directionality. The obtained channel parameters, such as DS and KF, serve as vital inputs to IIoT channel models. For MIMO channel measurements, the Tx antenna consists of a 4$\times$4 dual-polarized Uniform Planar Array (UPA) and the Rx antenna is a 4$\times$8 dual-polarized Uniform Cylindrical Array (UCA). Both Tx and Rx antenna arrays work at 5.15--5.85~GHz. Antenna elements are spaced with an interval of half the wavelength at 5.5~GHz. Rayleigh distances of Tx and Rx antenna arrays are 0.4912~m and 0.8006~m, satisfying the far-field assumption. The Tx and Rx antenna gains are 8.8~dBi and 8~dBi, respectively. The front-to-back gain, axial XPR (@0$^\circ$), and sector XPR (@$\pm 60^\circ$) of each antenna element at both sides are 17~dB, 15~dB, and 10~dB, respectively.
	
	\subsection{Measurement Cases}
	In order to explore comprehensive channel characteristics at Wi-Fi frequency bands in IIoT scenarios, both SISO and polarized MIMO channel measurements are conducted. The carrier frequency is set as 5.5 GHz, which is an unlicensed Wi-Fi band. Considering the obvious multipath effect in IIoT scenarios and capability of the channel sounder, the measurement bandwidth is set as 320 MHz to guarantee the high delay resolution. Two typical communication scenarios involving interactions between APs and terminals are considered. The first scenario focuses on wireless communications between an AP and the CPE installed on AGVs. The measurement route aligns with the AGV’s actual moving trajectories to capture realistic channel conditions. The second scenario is the communication link between the AP and various terminals scattered around the manufacturing area. Therefore, measurement points are distributed across the manufacturing area. During channel measurements, the Tx antenna is used to simulate the AP, while the Rx is used to imitate industrial terminals. Totally, there are three measurement cases and they provide a comprehensive evaluation of the wireless network at Wi-Fi bands in IIoT scenarios.
	\subsubsection{Case~1}
	To explore channel properties of communications between APs and the CPE mounted on AGVs, SISO channel measurements are carried out. The Rx antenna is set at a height of 0.3~m to simulate actual AGV conditions, while the Tx antenna is positioned at a height of 4.8~m, similar to the height of an actual AP. Measurement positions are distributed along two separate routes, each with a length of 90~m. The distance between adjacent Rx positions is 1~m, resulting in a total of 180 Rx positions. To analyze the power fluctuations when varying the Rx positions across several wavelength ranges, there are 7 locally amplified positions, designated as RxA1-1 to RxA1-7 in Fig.~\ref{IIoT_measurement_case}. The Rx antenna is moved with step size of 0.2~m (approximately four wavelengths at 5.5~GHz) to form a 5$\times$10 UPA, enabling the study of amplitude distribution for SSF.
	\subsubsection{Case~2}
	SISO channel measurements are performed to investigate wireless channel characteristics between APs and terminals suited in the manufacturing area. Heights of Tx and Rx antennas are set as 4.8~m and 1~m, respectively. The Tx position in Case~2 remains consistent with that in Case~1, i.e., Tx1 in Fig.~\ref{IIoT_measurement_case}. A total of 188 Rx positions are distributed around the manufacturing area, with a spacing of 1~m between adjacent Rx positions. The Three-Dimensional (3D) distances of all positions are 6--48.1~m in Case~1, 4.6--40.9~m in Case~2, and 3.1--39.7~m in office scenarios. Based on the antenna height and antenna beamwidth, the covered 3D distance between the Tx and Rx is larger than 5.1~m. Hence, we ignore these positions with 3D distances smaller than 5.1~m in the analysis. Similarly, there are 8 locally amplified positions, indicated as RxA2-1 to RxA2-8 in Fig.~\ref{IIoT_measurement_case}. 
	\subsubsection{Case~3}
	To study the angular and polarized characteristics of wireless channels between APs and terminals around the manufacturing area, 32$\times$64 polarized MIMO channel measurements are conducted. Antenna heights of the Tx and Rx are 4.5~m and 1.35~m, respectively. There are two Tx positions, with Tx2 located adjacent to a wall and 50 LOS positions distributed on one side of the Tx position. Meanwhile, Tx3 is situated at the center of an operation island and surrounded by 42 LOS positions. To distinguish position numbers in Case~1 and Case~2, the numbering of 92 positions in Case~3 is designated as Rx1'--Rx92'. In addition, the local coordinate systems of Tx and Rx antennas are provided in Fig.~\ref{IIoT_measurement_case}. Three measurement cases are summarized in Table~\ref{tab_three_cases}.
	\begin{table}[!t]
		\begin{center}	
			\caption{Configurations for three measurement cases.}
			\label{tab_three_cases}
			\setlength{\tabcolsep}{3.5pt}
			\begin{tabular}{|c|c|c|c|c|}
				\hline
				\textbf{Cases}& \textbf{Scenarios} & \textbf{Antenna Type}&\textbf{\makecell[c]{Antenna\\ Height (m)}}&\textbf{\makecell[c]{Measurement \\Positions}} \\
				\hline
				1& \makecell[c]{AGV's \\trajectory}& \makecell[c]{Omnidirectional\\ antennas (SISO) }& \makecell[c]{Tx: 4.8\\ Rx: 0.3}&\makecell[c]{Tx: 1\\ Rx: 180, RxA: 7} \\
				\hline
				2& \makecell[c]{Manufacturing\\ area}& \makecell[c]{Omnidirectional\\antennas (SISO)}& \makecell[c]{Tx: 4.8\\ Rx: 1}&\makecell[c]{Tx: 1\\ Rx: 188, RxA: 8} \\
				\hline
				3& \makecell[c]{Manufacturing\\ area}& \makecell[c]{Tx: UPA\\ Rx: UCA}& \makecell[c]{Tx: 4.5\\ Rx: 1.35}&\makecell[c]{Tx: 2\\ Rx: 92} \\
				\hline
			\end{tabular}
		\end{center}
	\end{table}
	
	\section{Measurement Data Processing}
	\label{sec:measurement data processing}
	During the measurement data processing, the Channel Impulse Response (CIR) is firstly obtained by calibrating out effects of the system response from the raw measurement data. For SISO channel measurements, the peak search algorithm is used to extract MultiPath Components (MPCs) from the delay PSD with a threshold determined by the maximum power and noise floor. Then, the received power is estimated by summing the power of all MPCs’ path gains. More details can be found in~\cite{Zhang2023}. For MIMO channel measurements, SMCs and DMCs are estimated to describe the wireless channel. The channel parameter estimation is based on the Maximum Likelihood (ML) principle. For SMCs, the state-of-the-art dual-polarized Space-Alternating Generalized Expectation-maximization (SAGE) algorithm~\cite{Yin2003} is employed. Channel parameters of each path can be extracted, including the amplitude, delay, Elevation angle Of Arrival (EoA), Azimuth angle Of Arrival (AoA), Elevation angle Of Departure (EoD), and Azimuth angle Of Departure (AoD). Regarding DMCs, their parameters in the delay PSD and angular PSD are estimated by a Levenberg-Marquardt algorithm with Fisher scoring. Traditionally, DMCs in existing channel measurements are estimated by a single DMC process within the delay PSD model. However, measurement results in IIoT scenarios reveal that there are multiple independent DMC processes. Therefore, the approach in~\cite{Ribeiro2007} is extended to accommodate and effectively process such multi-process measurement data in IIoT scenarios.
	\subsection{SMCs}
	Let $\textbf{H}_\text{meas}(\boldsymbol{f}) \in \mathbb{C}^{ M_\text{R} \times M_\text{T} \times M_f}$ denote the measured channel transfer function matrix, which is obtained by Fast Fourier Transform (FFT) of CIR $\boldsymbol{h}_\text{meas}(\boldsymbol{\tau}) \in \mathbb{C}^{ M_\text{R} \times M_\text{T} \times M_f}$. Here, $M_\text{R} = 64$, $M_\text{T} = 32$, and $M_f$ are the number of Rx antennas, Tx antennas, and frequency points, respectively. Vectors of frequency points and delay indices are denoted by $\boldsymbol{f} \in \mathbb{R}^{M_f}$ and $\boldsymbol{\tau} \in \mathbb{R}^{M_f}$, respectively. The complex and real domains are represented by $\mathbb{C}$ and $\mathbb{R}$. SMCs are characterized by the parameter set ${\boldsymbol{\Theta}_\text{S}}=\left[\boldsymbol{\theta}_1,\boldsymbol{\theta}_2,...,\boldsymbol{\theta}_L\right]$, where $\boldsymbol{\theta}_l = [\theta_l^\text{R}; \phi_l^\text{R}; \theta_l^\text{T}; \phi_l^\text{T}; \tau_l; \text{vec}(\boldsymbol{\alpha}_l)]$. Here, symbols $\theta_l^\text{R}$, $\phi_l^\text{R}$, $ \theta_l^\text{T}$, $\phi_l^\text{T}$, and $\tau_l$ denote the EoA, AoA, EoD, AoD, and propagation delay, respectively. Besides, $\boldsymbol{\alpha}_l=\left[\alpha_l^\text{VV},\alpha_l^\text{VH};\alpha_l^\text{HV},\alpha_l^\text{HH}\right] \in \mathbb{C}^{2\times 2}$ is the complex amplitude matrix of four polarization modes, and $\text{vec}(\cdot)$ denotes the vectorization operation. The estimated dual-polarized transfer function matrix of SMCs can be characterized by the superposition of $L$ individual paths as  
	\begin{equation}	
		\label{sage}	
		\hat{\textbf{H}}_\text{S}(\boldsymbol{f};\boldsymbol{\Theta}_\text{S})= \sum_{l=1}^{L} \textbf F_\text{R}(\theta_l^\text{R}, \phi_l^\text{R}) \boldsymbol{\alpha}_l \textbf F_\text{T}(\theta_l^\text{T}, \phi_l^\text{T})^Te^{j2\pi\boldsymbol{f}\tau_l} .
	\end{equation}
	
	In~(\ref{sage}), radiation patterns of Rx and Tx antenna arrays are represented by $\textbf F_\text{R}(\theta_l^\text{R}, \phi_l^\text{R}) \in \mathbb{C}^{M_\text{R}\times 2}$ and $\textbf F_\text{T}(\theta_l^\text{T}, \phi_l^\text{T}) \in \mathbb{C}^{M_\text{T}\times 2}$, respectively. The transpose operation is denoted by $(\cdot)^T$. A comprehensive description of the initialization and iteration process can be found in~\cite{Yin2003}. In a word, the estimated parameters $\hat{\boldsymbol{\Theta}}_\text{S}$ maximize the likelihood function
	\begin{equation}	
		\label{Likelihood_SC}	
		\hat{\boldsymbol{\Theta}}_\text{S} = \mathop{\arg\max}_{\boldsymbol{\Theta}_\text{S}} \left \{-\bigg \lVert\text{vec}\Big \{\textbf{H}_\text{meas}(\boldsymbol{f})\Big \} - \text{vec}\left\{\hat{\textbf{H}}_\text{S}(\boldsymbol{f};\boldsymbol{\Theta}_\text{S})\right\}\bigg\rVert^2_\text{F}\right\}
	\end{equation}
	where $\vert|\cdot\vert|_\text{F}$ is the Frobenius norm.
	
	\subsection{DMCs}
	The measured channel transfer function matrix of DMCs ${\textbf{H}}_{\text{D,meas}}(\boldsymbol{f}) \in \mathbb{C}^{ M_\text{R} \times M_\text{T} \times M_f}$ is calculated by subtracting SMCs from the measurement data as
	\begin{equation}	
		\label{DMC}	
		{\textbf{H}}_{\text{D,meas}}(\boldsymbol{f}) = \textbf{H}_\text{meas}(\boldsymbol{f}) - \hat{\textbf{H}}_\text{S}(\boldsymbol{f};\boldsymbol{\Theta}_\text{S}).
	\end{equation}
	The measured covariance matrix is represented by
	\begin{equation}
		\label{covariance matrix of meas}
		\textbf{R}_\text{D,meas} = \text{vec}\big({\textbf{H}}_{\text{D,meas}}(\boldsymbol{f})\big)\text{vec}\big({\textbf{H}}_{\text{D,meas}}(\boldsymbol{f})\big)^H	
	\end{equation} 
	where $(\cdot)^H$ is the Hermitian transpose. The estimation of DMCs is based on correlation matrices in frequency and spatial domains. The delay PSD is characterized by an exponential decay distribution as~\cite{Ribeiro2007} 
	\begin{eqnarray}
		\label{PDP_DMC}
		\boldsymbol{\Psi}({\tau}) = \left\{
		\begin{array}{rcl}
			& \alpha_0, & \tau < \tau_d \\ 
			& \alpha_1/2, & \tau = \tau_d\\
			& \alpha_1 e^{-B_d(\tau-\tau_d)}, & \tau > \tau_d
		\end{array} 
		\right.
	\end{eqnarray}
	where $B_d$, $\alpha_1$, $\alpha_0$, and $\tau_d$ are the power decay rate, maximum power, minimum power, and base delay related to the Tx--Rx distance, respectively. The parameter vectors describing correlation matrices of $K$ uncorrelated DMC processes in the delay domain are denoted as $\boldsymbol{\theta}_\text{F}^k = \left[\alpha_1^k,B_d^k,\tau_d^k \right]$. Hence, the covariance matrix in the frequency domain is given by the sum of covariance matrices of $K$ independent DMC processes and white noise as
	\begin{equation}
		\hat{\textbf{R}}(\boldsymbol{\theta}_{\text{F}}) = \alpha_0 \textbf{I} + \text{toep}\left(\boldsymbol{\kappa}\left(\boldsymbol{\theta}_\text{F}\right),\boldsymbol{\kappa}\left(\boldsymbol{\theta}_\text{F}\right)^H  \right)\in \mathbb{C}^{M_f\times M_f}
	\end{equation}
	where $\textbf{I}$ is an identity matrix and $\text{toep}(\cdot)$ is the Toeplitz function. Besides, the sample of frequency correlation matrix is writen as $\boldsymbol{\kappa}\left(\boldsymbol{\theta}_\text{F}\right)=\sum_{k=1}^{K}\boldsymbol{\kappa}\left(\boldsymbol{\theta}_\text{F}^k\right)$ and the $k$-th component is calculated similar to~\cite{Ribeiro2007}.
	
	The joint azimuth--elevation angular PSD is fitted by the multimodal Von Mises-Fisher (VMF) distribution and an example at the Rx side is given by
	\begin{equation}
		\label{vons_pdf}
		f(\vartheta,\varphi;\boldsymbol{\theta}_\text{A}^\text{R}) = \sum_{q=1}^{Q} \epsilon_q^\text{R} \frac{{(\kappa_q^\text{R})^\frac{1}{2}}}{(2\pi)^\frac{3}{2}I_{\frac{1}{2}}(\kappa_q^\text{R})}e^{\kappa_q^\text{R}\boldsymbol{\mu}_q^\text{R}\boldsymbol{\Omega}(\vartheta,\varphi)}
	\end{equation}
	where angular parameters at the Rx side are represented by $\boldsymbol{\theta}_\text{A}^\text{R}=\left[\boldsymbol{\theta}_{\text{A},1}^\text{R};...;\boldsymbol{\theta}_{\text{A},Q}^\text{R} \right]$ and $\boldsymbol{\theta}_{\text{A},q}^\text{R} = \left[ \vartheta_{q}^\text{R}, \varphi_{q}^\text{R}, \kappa_{q}^\text{R}, \epsilon_{q}^\text{R}\right]$. Here, $Q$ is the number of components and $\epsilon_q^\text{R}$ is the proportion of each component with $\sum_{q=1}^{Q}\epsilon_q^\text{R}=1$. Besides, $\kappa_q^\text{R} \geq 0$ is the concentrated parameter of DMCs and $\boldsymbol{\mu}_q^\text{R}=[\cos(\vartheta_{q}^\text{R})\cos(\varphi_{q}^\text{R}),\cos(\vartheta_{q}^\text{R})\sin(\varphi_{q}^\text{R}),\sin(\vartheta_{q}^\text{R})]$ is the direction of arrival angle determined by the AoA $\varphi_{q}^\text{R}$ and EoA $\vartheta_{q}^\text{R}$. Symbol $I_{\frac{1}{2}}(\cdot)$ is the modified Bessel function of the first kind and the spherical unit vector is denoted by $\boldsymbol{\Omega}(\vartheta,\varphi)=[\cos(\vartheta)\cos(\varphi),\cos(\vartheta)\sin(\varphi),\sin(\vartheta)]^T$. Under the condition of $\kappa_q^\text{R}=0$, the angular PSD reduces to a uniform distribution.
	The angular PSD and parameters at the Tx side are similar to those at the Rx side and will not be explained in detail. 
	
	It is assumed that the covariance matrix of DMCs can be factorized into a Kronecker product. Hnece, the estimated covariance matrix of DMCs is represented by
	\begin{equation}
		\label{covariance matrix of estimation}
		\hat{\textbf{R}}(\boldsymbol{\Theta}_\text{D}) = \hat{\textbf{R}}(\boldsymbol{\theta}_{\text{F}}) \otimes \hat{\textbf{R}}(\boldsymbol{\theta}_{\text{A}}^\text{R}) \otimes \hat{\textbf{R}}(\boldsymbol{\theta}_{\text{A}}^\text{T})	
	\end{equation} 
	where $\otimes$ is the Kronecker product. Besides,  $\hat{\textbf{R}}(\boldsymbol{\theta}_{\text{A}}^\text{T}) \in \mathbb{C}^{M_\text{T}\times M_\text{T}}$ and $\hat{\textbf{R}}(\boldsymbol{\theta}_{\text{A}}^\text{R}) \in \mathbb{C}^{M_\text{R}\times M_\text{R}}$ indicate the estimated covariance matrices in the angular domain at both Tx and Rx sides. The optimization of DMC parameters ${\boldsymbol{\Theta}}_\text{D} = \left[\boldsymbol{\theta}_\text{F};\boldsymbol{\theta}_{\text{A}}^\text{R};\boldsymbol{\theta}_{\text{A}}^\text{T}\right]$ can be performed by maximizing the log-likelihood function as~\cite{Hanssens2018-2}  
	\begin{equation}	
		\begin{small}
			\label{Likelihood_DMC}	
			\hat{\boldsymbol{\Theta}}_\text{D} = \mathop{\arg\max}_{\boldsymbol{\Theta}_\text{D}} \bigg\{-\ln\left(\det\left(\hat{\textbf{R}}\left(\boldsymbol{\Theta}_\text{D}\right)\right)\right) - \text{tr}\left(\hat{\textbf{R}}(\boldsymbol{\Theta}_\text{D})^{-1}\textbf{R}_\text{D,meas}\right)\bigg\}
		\end{small}
	\end{equation}
	where $\det\left(\cdot\right)$, $\text{tr}\left(\cdot\right)$, and $\left(\cdot\right)^{-1}$ denote the determinant, trace, and inverse of a matrix, respectively. Since parameters in delay and angular domains are assumed to be independent of each other, they are estimated separately to reduce the complexity. 
	
	To solve the ML problem, an initial solution is first provided and it significantly impacts the algorithm's convergence efficiency. For delay parameters, identifying the base delay associated with each DMC process is important. By observing the measured delay PSD, we can find that there exist multiple independent DMC processes. To detect the transition points in the slope of each process, the first-order difference technique is applied, thereby determining the dataset of each DMC process with different data lengths. Then, the initialization approach in~\cite{Ribeiro2007} is adopted to estimate initial channel parameters of each dataset. Given that each DMC process spans a unique duration, it is crucial to normalize the base delays so that they can be computed under an equivalent number of frequency points. As for the angular parameters, an initial approximation is derived from the Bartlett spectrum. Subsequently, the Levenberg-Marquardt algorithm with gradient function and fisher information matrix is employed to get iterative results. Detailed calculations of these functions can be found in~\cite{Ribeiro2007}. Using the estimated SMC and DMC parameters, the CIR can be synthetically reconstructed and compared to the measurement result. Firstly, a matrix of zero-mean complex Gaussian distributed values $\boldsymbol{z} \in \mathbb{C}^{M_f \times M_\text{R}M_\text{T}}$ is introduced. Then, the Cholesky decomposition is performed to obtain the transform matrix $\boldsymbol{L}_\text{F}$ with $\hat{\textbf{R}}(\boldsymbol{\theta}_{\text{F}}) = \boldsymbol{L}_\text{F}\boldsymbol{L}_\text{F}^H$ in the delay-frequency domain and $\boldsymbol{L}_\text{A}$ with $\hat{\textbf{R}}(\boldsymbol{\theta}_{\text{A}}) = \boldsymbol{L}_\text{A}\boldsymbol{L}_\text{A}^H$ in the angular domain. Finally, transfer function matrices of the estimated DMCs and full channel can be obtained by
	\begin{equation}
		\hat{\textbf{H}}_\text{D}(\boldsymbol{f};\boldsymbol{\Theta}_\text{D})=\boldsymbol{L}_\text{F} \boldsymbol{z} \boldsymbol{L}_\text{A}
	\end{equation}
	\begin{equation}
		\hat{\textbf{H}}(\boldsymbol{f}) = \hat{\textbf{H}}_\text{S}(\boldsymbol{f};\boldsymbol{\Theta}_\text{S}) + \hat{\textbf{H}}_\text{D}(\boldsymbol{f};\boldsymbol{\Theta}_\text{D}).
	\end{equation}
	
\begin{figure}[!t]
		\centering
		\begin{minipage}[!t]{\columnwidth}
			\centering
			\subfigure[]{\includegraphics[width = 0.49\columnwidth]{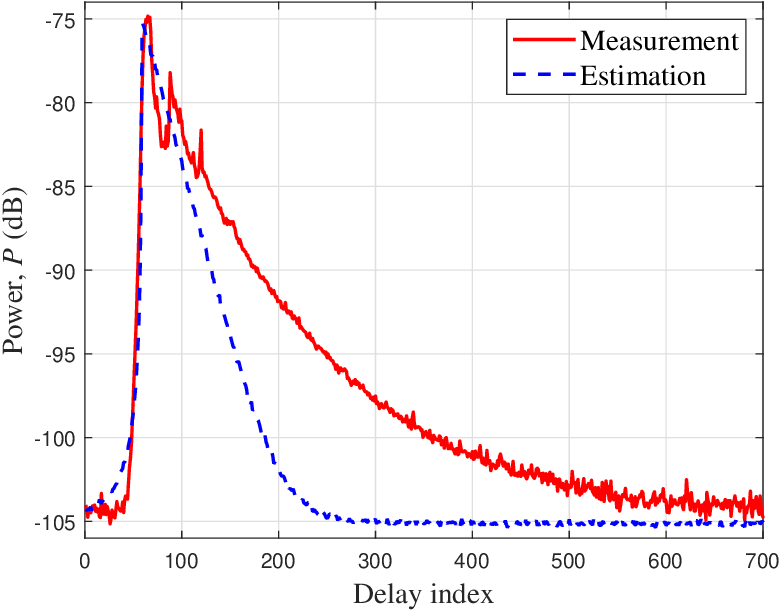}}
			\subfigure[]{\includegraphics[width = 0.49\columnwidth]{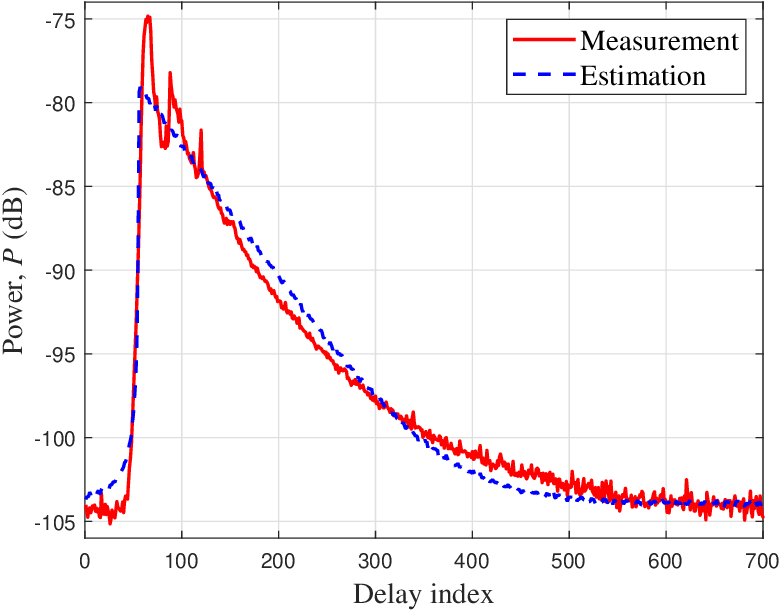}}
		\end{minipage}	
		\begin{minipage}[!t]{\columnwidth}
			\centering
			\subfigure[]{\includegraphics[width = 0.49\columnwidth]{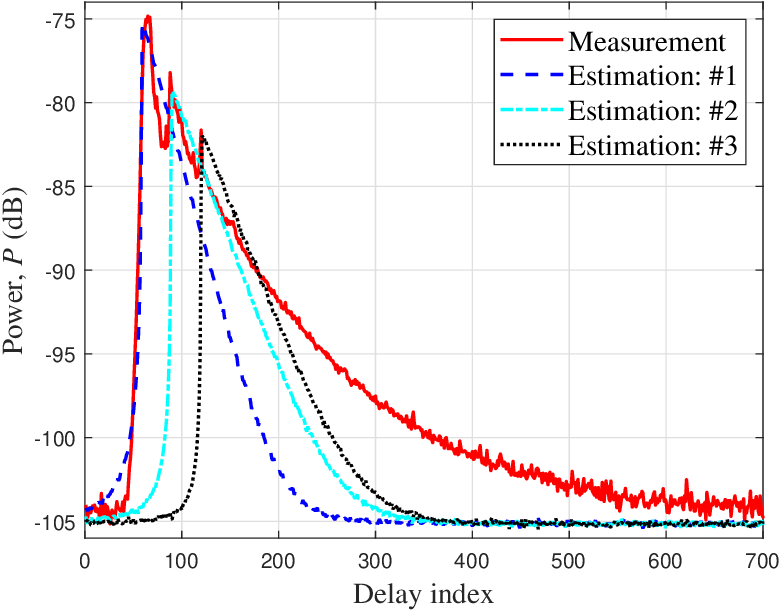}}
			\subfigure[]{\includegraphics[width = 0.49\columnwidth]{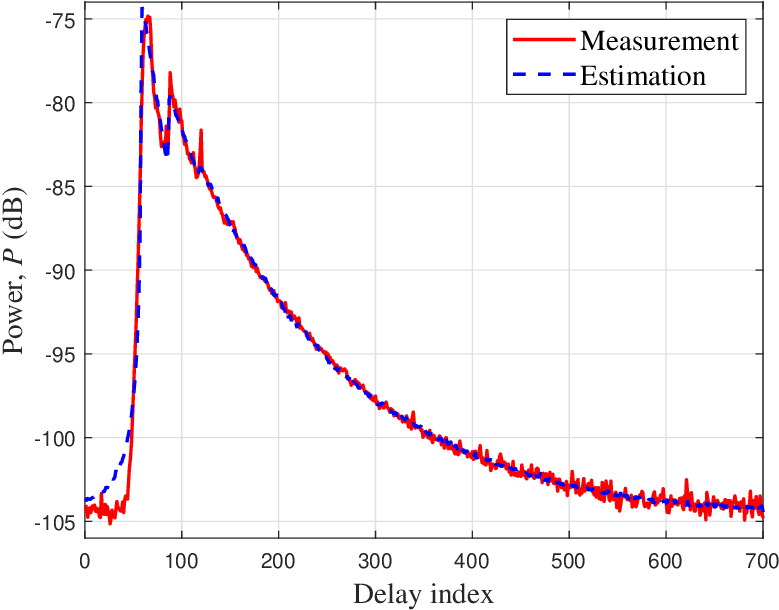}}
		\end{minipage}	
		\caption{Comparisons of (a) initial and (b) iterative estimation results for one DMC process as well as (c) initial and (d) iterative estimation results for three DMC processes.}
		\label{DMC_freq_est}
	\end{figure}

	Fig.~\ref{DMC_freq_est} compares estimation and measurement results of a single DMC process and those of three DMC processes during both initialization and iteration phases. For a single DMC process, there is an obvious difference in power decay rates between the initial estimate and iterated solution, indicating the presence of at least two distinct slopes. The iterative result captures about 79\% of the total power. In the proposed method, three independent DMC processes are modeled in the delay PSD. Fig.~\ref{DMC_freq_est}(c) shows the initialization results based on this assumption. After an iterative process with ten iterations, the estimated delay PSD aligns closely with the measurement data, capturing about 99.6\% of the total power, thereby demonstrating a satisfactory match.  
	
	\section{Measurement Results and Analysis}
	\label{sec:Results and Analysis}
	In this section, we present a comprehensive analysis of channel characteristics for SISO and MIMO channel measurements in IIoT scenarios. The obtained LSPs such as DS, KF, and ASs, play an important role in generating accurate SSF channel models. By illustrating the effects of DMCs on SVs and channel capacities, we highlight the critical importance of accurately incorporating DMC characteristics into IIoT channel models. Therefore, the comparative analysis of channel characteristics between SMCs and DMCs provide a basis for the development of DMC models.
	
	\subsection{SISO Measurement Results and Analysis: Delay PSD, PL, SF, ED, DS, KF, and Amplitude Distribution of SSF}	
	\subsubsection{Delay PSD}
\begin{figure}[!t]
	\centering
	\begin{minipage}[!t]{\columnwidth}
		\centering
		\subfigure[]{\includegraphics[width = 0.72\columnwidth]{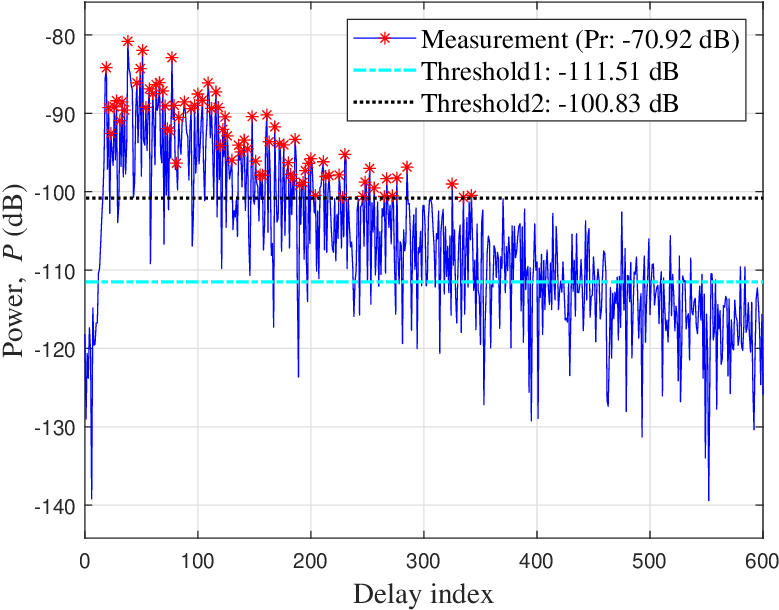}}
		\subfigure[]{\includegraphics[width = 0.72\columnwidth]{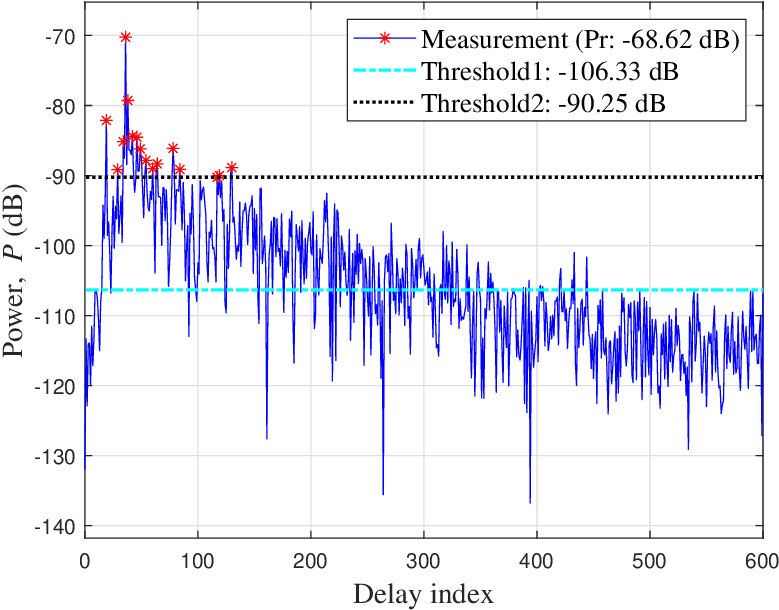}}
	\end{minipage}
	\caption{Measured delay PSDs in IIoT scenarios at (a) position Rx71 of Case~1 and (b) position Rx281 of Case~2.}
	\label{delay_PSD_SISO}
	\end{figure}

	Fig.~\ref{delay_PSD_SISO} displays the measured delay PSDs of positions Rx71 in Case~1 and Rx281 in Case~2, which are located at identical positions with different heights of Rx antennas. During channel measurements, the Tx antenna is at the right rear of the metal pillar and the link between the Tx and Rx is obstructed by the metal pillar. Hence, the LOS path is not dominant and they are considered as NLOS positions. The estimated MPCs, indicated by `*' in Fig.~\ref{delay_PSD_SISO} are distinguished by applying a threshold. The threshold is determined by the larger value of 10 dB above the noise floor (Threshold1) and 20 dB below the maximum power (Threshold2). It is evident that a reflection path with stronger power exists in Case~2, thereby resulting in a higher threshold for determining MPCs in Case~2 compared to Case~1. This leads to the extraction of more MPCs in Case~1 and similar received power (denoted as Pr in Fig.~\ref{delay_PSD_SISO}) as in Case~2. Furthermore, the delay PSDs exhibit abundant MPCs and evident trailing in IIoT scenarios. For SISO channel measurements, it is difficult to distinguish SMCs and DMCs from the delay PSD without the angular information. Channel characteristics are analyzed based on the all extracted paths, thereby including SMCs and DMCs. Comparisons between SMCs and DMCs are demonstrated through MIMO channel measurements.

	\subsubsection{PL and SF}
	 The PL helps to conduct the link budget, determine the numbers and locations of required APs, and assess the intra-network interference. There are two empirical PL models available for fitting measurement results, i.e., Close-In (CI) PL model and Floating Intercept (FI) PL model~\cite{Jiang2020}. The equations of these models can be expressed by
	\begin{equation}\label{CI}
		PL = 32.4 + 20 \text{log}_{10}(f_c) + 10n \text{log}_{10}(d) + X_\sigma
	\end{equation}
	\begin{equation}\label{FI}
		PL = \beta + 10n \text{log}_{10}(d) + X_\sigma.
	\end{equation}	
	The center frequency and distance between Tx and Rx are denoted as $f_c$ and $d$, respectively. The fitted parameters include the Path Loss Exponent (PLE) $n$, the intercept $\beta$ in dB, and the SF $X_\sigma$ assumed to follow a normal distribution with a mean of zero and a standard variance of $\sigma$ in dB. If $n=2$ in eq.~\ref{CI}, the PL of free space is obtained. To keep a fair comparison between the CI and FI models, the intercept of the CI model is designated by $32.4 + 20 \text{log}_{10}(f_c)$.
	
\begin{figure}[!t]
		\centering
		\begin{minipage}[!t]{\columnwidth}
			\centering
			\subfigure[]{\includegraphics[width = 0.72\columnwidth]{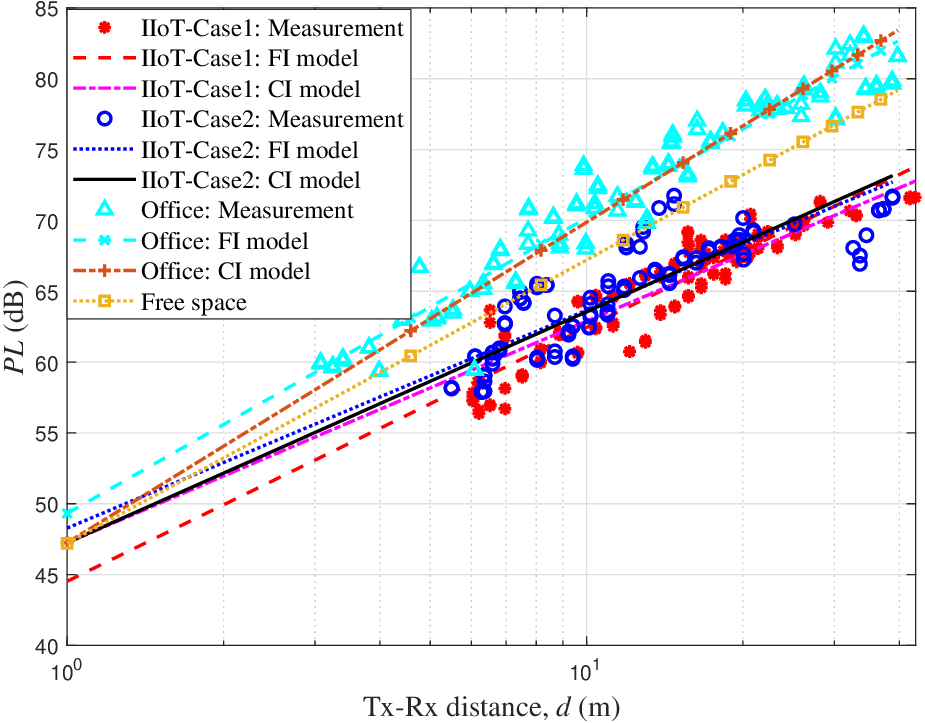}}
			\subfigure[]{\includegraphics[width = 0.72\columnwidth]{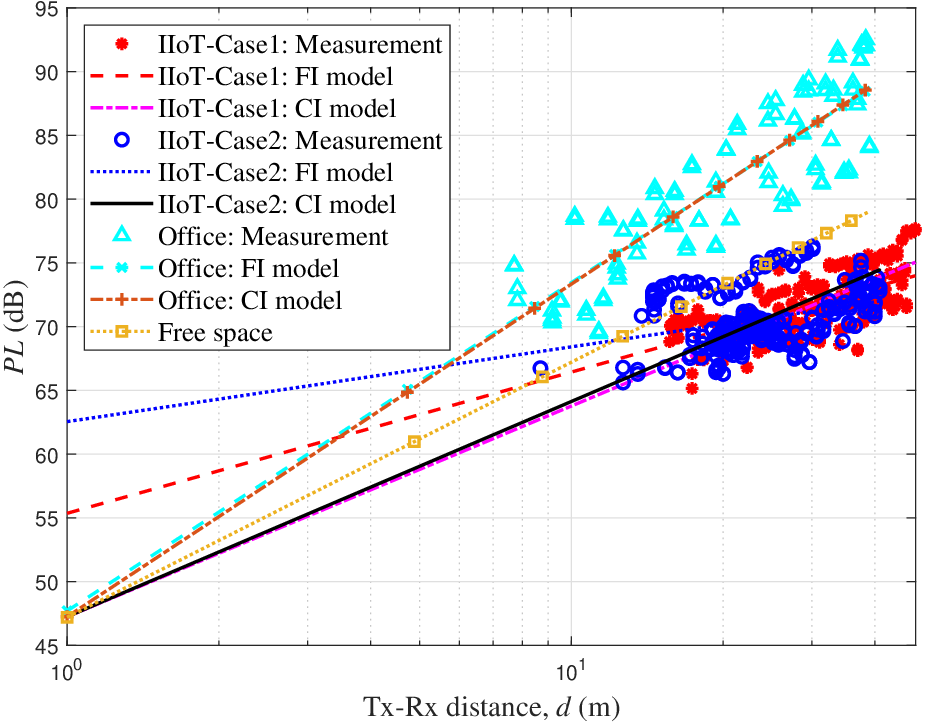}}
		\end{minipage}		
		\caption{PL of measurement, CI model, and FI model results for Case~1, Case~2, and office scenarios in (a) LOS scenarios and (b) NLOS scenarios.}
		\label{PL}
	\end{figure}

	Fig.~\ref{PL} demonstrates the PL of measurement and model results in Case~1, Case~2, and office environments. The PL of IIoT scenarios is lower than those of the office scenario and free space. Due to the larger physical size and more metal objects in factory environments, there are rich scattering MPCs that can increase the received power. In IIoT scenarios, the PL results in Case~1 and Case~2 exhibit minimal disparity. In Case~1, the Rx antenna height is merely 0.3~m, which is significantly lower than most scatterers. Hence, the number of MPCs that satisfy the threshold is larger and the received power is high. In Case~2, there exists a higher probability of receiving wall reflection paths with strong power compared to that in Case~1. Consequently, there is negligible difference in PL between both cases. The PL of Case~2 in NLOS scenarios has two segments with the second segment being 5--10 dB higher than the first. This increase in PL is attributed to locations Rx340--368 where the height of Tx antenna is slightly lower than that of nearby scatterers on the operation island and the blockage loss is larger.
	
	Parameters of the two fitted PL models and 3GPP TR 38.901~\cite{3GPP} (referred to as 3GPP in the subsequent context) model are summarized in Table~\ref{tabPL}. Notably, the PLE values for IIoT scenarios are smaller than those of 3GPP model, which aligns with findings from~\cite{Wang2021,Jiang2020,Schmieder2019,Ai2015}. The difference between CI and FI models is slight in LOS scenarios but due to the presence of two segments in NLOS scenarios of Case~2, the parameters of FI model differ from those of the CI model. When excluding the data from the second segment, the three parameters $(n, \beta,\sigma)$ for CI and FI models are (1.61, 47.25, 1.33) and (1.15, 53.7, 1.2), respectively.	
	\begin{table}
		\begin{center}
			\caption{Parameters of CI and FI models in LOS and NLOS scenarios.}
			\setlength{\tabcolsep}{5.2pt}
			\begin{tabular}{|c|c|c|c|c|c|c|c|}
				\hline				
				\multicolumn{2}{|c|}{\textbf{Cases}}&\multicolumn{3}{c|}{\textbf{LOS scenario}}& \multicolumn{3}{c|}{\textbf{NLOS scenario}} \\
				\cline{1-8}
				\multicolumn{2}{|c|}{Parameters}&\makecell[c]{CI\\ model}&\makecell[c]{FI \\model}&\makecell[c]{3GPP \\model}&\makecell[c]{CI\\ model}&\makecell[c]{FI \\model}&\makecell[c]{3GPP \\model}\\
				\hline
				\multirow{3}{*}{IIoT-Case~1}&$n$& 1.56& 1.79& 2.15& 1.65& 1.11& 3.57\\
				&$\beta$& 47.25& 44.53& 45.91& 47.25& 55.36& 33.41\\
				&$\sigma$& 1.6& 1.52& 4& 1.91& 1.76& 5.7\\	
				\hline
				\multirow{3}{*}{IIoT-Case~2}&$n$& 1.63& 1.54& 2.15& 1.69& 0.59& 3.57\\
				&$\beta$& 47.25& 48.28& 45.91& 47.25& 62.55& 33.41\\
				&$\sigma$& 2.12& 2.11& 4& 2.63& 2.23& 5.7\\
				
				\hline
				\multirow{3}{*}{Office}&$n$& 2.26& 2.09& 1.73& 2.61& 2.57& 3.83\\
				&$\beta$& 47.25& 49.31& 47.25& 47.25& 47.71& 47.25\\
				&$\sigma$& 1.85& 1.76& 3& 3.13& 3.13& 8.03\\			
				\hline	
			\end{tabular}
			\label{tabPL}
		\end{center}
	\end{table}		
	\subsubsection{ED and DS}
	ED and DS characterize the channel dispersion in the time domain. DS determines the guard interval length in orthogonal frequency-division multiplexing systems and tap lengths in equalizer designs when combating inter-symbol interference~\cite{Zhong2020}. Among the MPCs that satisfy the threshold, ED refers to the delay difference between the last MPC and the first MPC, i.e., $ED = \tau_L - \tau_1$. The DS can be calculated as~\cite{Zheng2023-2}
	\begin{equation}		DS=\sqrt{\frac{\sum^L_{l=1}P_{l}\tau_{l}^2}{\sum^L_{l=1}P_{l}}-\left(\frac{\sum^L_{l=1}P_{l}\tau_{l}}{\sum^L_{l=1}P_{l}}\right)^2}
		\label{DS}
	\end{equation}
	where $\tau_l$ and $P_l$ are the delay and power of $l$th MPC, respectively. 
	
\begin{figure}[!t]
		\centering
		\includegraphics[width=0.72\columnwidth]{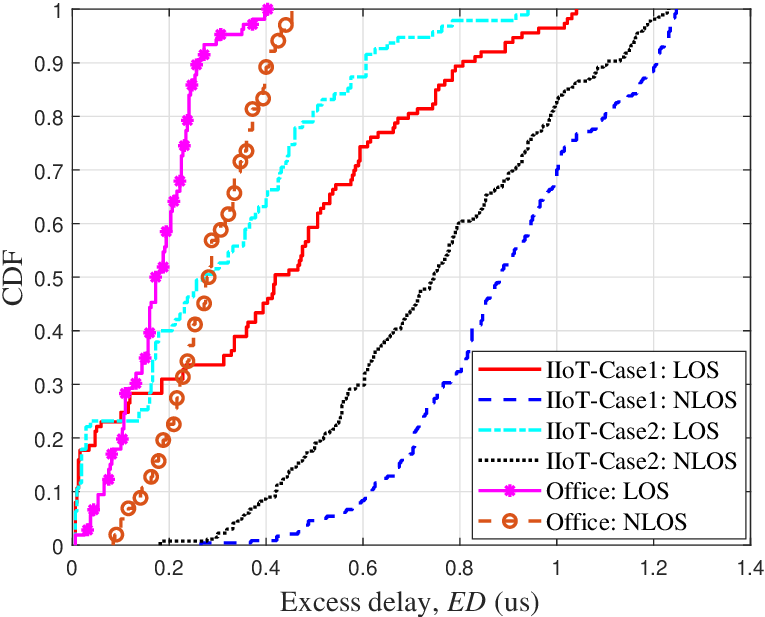}	
		\caption{CDFs of the measured ED in Case~1, Case~2, and office scenarios.}
		\label{ED}
	\end{figure}

	Fig.~\ref{ED} illustrates the Cumulative Distribution Functions (CDFs) of measured EDs in Case~1, Case~2, and office environments. It is evident that EDs are significantly lower in LOS scenarios compared to NLOS scenarios for both IIoT and office environments, due to the complex scattering environment in NLOS scenarios. Moreover, higher ED values are observed in Case~1 owing to the lower height of the receiving antenna and a larger number of MPCs. In addition, EDs in office environments are relatively smaller than those in IIoT scenarios where there are abundant metal objects. In~\cite{Zhong2020}, the size of the measurement area is 115$\times$28$\times10$~m³ and the maximum 3D distance is about 118~m, larger than that in this work. Values of the ED for both LOS and NLOS scenarios were found to be 0.9--1.4~us and 0.4--1.7~us, respectively. Corresponding EDs in~\cite{Razzaghpour2019} were within the intervals of 0--140~ns and 0--200~ns for LOS and NLOS conditions, respectively. They are much smaller than results in~\cite{Zhong2020} and this work since the maximum 3D distance is about 9~m. Therefore, larger measured 3D distance results in larger ED.
	
\begin{figure}[!t]
		\centering
		\includegraphics[width=0.7\columnwidth]{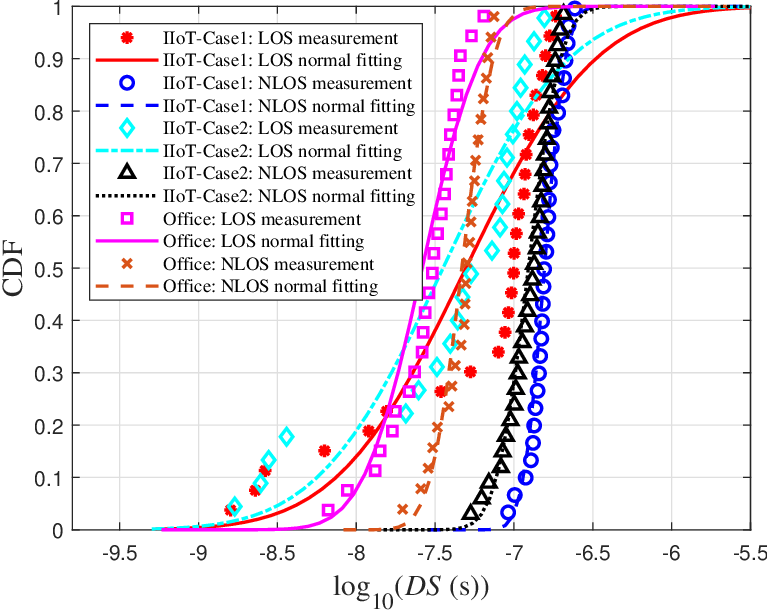}	
		\caption{DS of measurement and fitting results in Case~1, Case~2, and office scenarios.}
		\label{DS_SISO}
	\end{figure}

	Fig.~\ref{DS_SISO} presents the measurement and log-normal distribution fitting results of DS in Case~1, Case~2, and office environments, with the corresponding distribution parameters listed in \mbox{Table~\ref{tabDS_KF}}. The DS and ED exhibit a similar trend because of analogous reasons. However, certain values may deviate from the expected distribution when Tx--Rx distance is small since the dominant LOS path can result in smaller DSs. Therefore, standard deviations in Case~1 and Case~2 are large. The DS in Case~1 is larger than that in Case~2 because of richer MPCs. Compared with results in the literature, distribution parameters of DSs are different because of different measurement configurations and environments. However, authors in~\cite{Jiang2020} investigated the effect of antenna heights on DSs and revealed that a lower antenna height can result in larger DSs, which is consistent with our findings.  
	
	\begin{table}[!t]
		\begin{center}
			\caption{Parameters of DS and KF in LOS and NLOS scenarios.}
			\setlength{\tabcolsep}{6pt}
			\begin{tabular}{|c|c|c|c|c|}			
				\hline
				\textbf{Parameters}& \textbf{Scenarios}& \textbf{IIoT-Case~1}& \textbf{IIoT-Case~2}& \textbf{Office} \\
				\hline
				\multirow{2}{*}{$\mu_{\text{lg}(\text{DS(s)})}$}& LOS& -7.41& -7.53& -7.58\\
				\cline{2-5}
				& NLOS& -6.81& -6.91& -7.33\\
				\hline
				\multirow{2}{*}{$\sigma_{\text{lg}(\text{DS(s)})}$}& LOS& 0.76& 0.71& 0.29\\
				\cline{2-5}
				& NLOS& 0.1& 0.16& 0.16 \\
				\hline
				\multirow{2}{*}{$\mu_{\text{lg}(\text{KF})}$}& LOS& 2.12& 3.44& 0.57\\
				\cline{2-5}
				& NLOS& -4.96& -3.66& -2.12\\
				\hline
				\multirow{2}{*}{$\sigma_{\text{lg}(\text{KF})}$}& LOS& 7.67& 7.66& 4.77\\
				\cline{2-5}
				& NLOS& 3.54& 3.53& 3.16 \\
				\hline							
			\end{tabular}
			\label{tabDS_KF}
		\end{center}
	\end{table}	
	\subsubsection{KF}
\begin{figure}[!t]
		\centering
		\includegraphics[width=0.7\columnwidth]{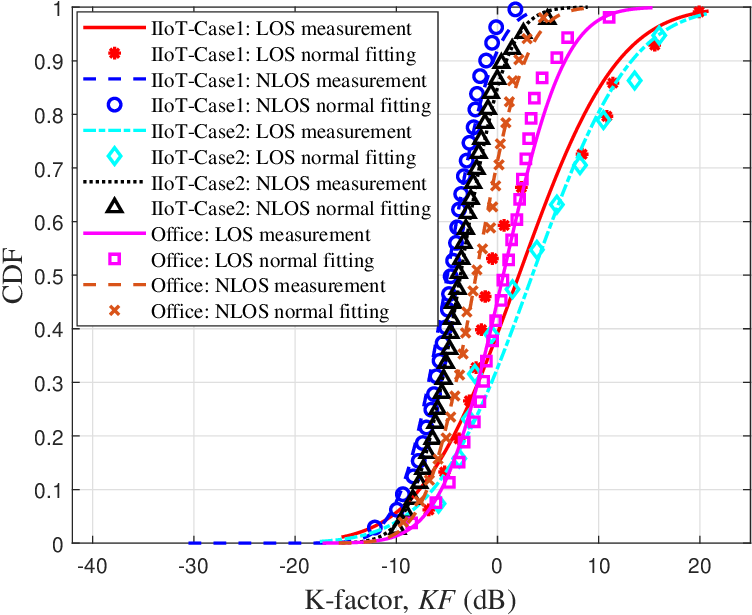}	
		\caption{KF of measurement and fitting results in Case~1, Case~2, and office scenarios.}
		\label{KF}
	\end{figure}

	The calculation of KF is consistent with the moment-based method in the frequency domain~\cite{Tang2019}. Fig.~\ref{KF} presents the measurement and normal fitting results and distribution parameters are provided in Table~\ref{tabDS_KF}. The KF in Case~1 is smaller than that in Case~2 due to the lower antenna height. Results in~\cite{Guan2020} also showed that higher antenna heights can decrease the probability of blockage and lead to a higher KF. In addition, the KF in IIoT scenarios is larger than that in office scenarios as some Rx positions are suited near to Tx without any obstructions between them. These positions have smaller DS and the strong LOS path leads to a higher KF.
	\subsubsection{Amplitude Distribution of SSF}
\begin{figure}[!t]
		\centering
		\begin{minipage}[!t]{\columnwidth}
			\centering
			\subfigure[]{\includegraphics[width = 0.49\columnwidth]{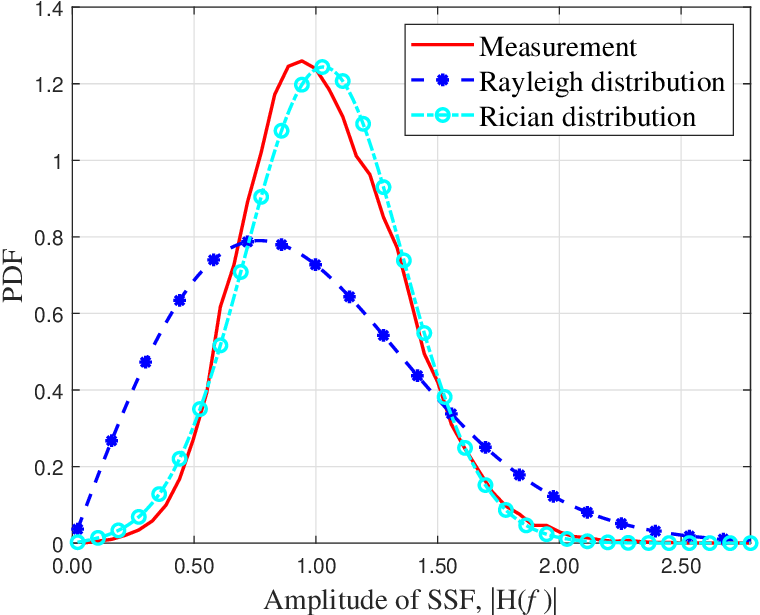}}
			\subfigure[]{\includegraphics[width = 0.49\columnwidth]{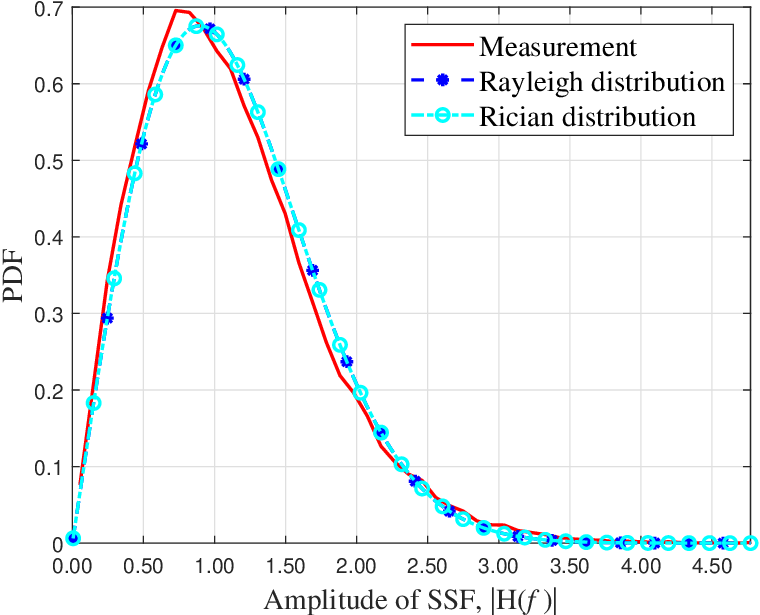}}
		\end{minipage}
		\caption{Amplitude distribution of SSF for measurement and fitting results in IIoT scenarios at (a) position RxA1-5 and (b) position RxA2-1.}
		\label{SSF}
	\end{figure}

	There are 7 and 8 locally amplified positions in Case~1 and Case~2, respectively. The amplitude distribution of all frequency points for each amplified position is analyzed across 50 local positions. Fig.~\ref{SSF}(a) is the amplitude distribution at position RxA1-5, fitted well a by Rician distribution instead of Rayleigh distribution. The LOS path is dominant at this position and the KF calculated based on distribution parameters is 6.45~dB. In addition, the amplitude distribution of position RxA1-4 also follows a Rician distribution and its KF is 3.82~dB. Fig.~\ref{SSF}(b) depict the well-fitted Rayleigh and Rician distributions at position \mbox{RxA2-1}. These two distributions coincide with each other due to a small KF of 0. Except positions RxA1-4 and RxA1-5, all other positions exhibit Rayleigh distributions in their amplitude distributions, consistent with results in~\cite{Tanghe2010} and~\cite{Karedal2004}. 	
	\subsection{MIMO Measurement Results and Analysis: Delay, Angular, Power, and Polarization Characteristics of SMCs and DMCs}
	\subsubsection{Delay Characteristics}
\begin{figure}[!t]
		\centering
		\begin{minipage}[!t]{\columnwidth}
			\centering
			\subfigure[]{\includegraphics[width = 0.7\columnwidth]{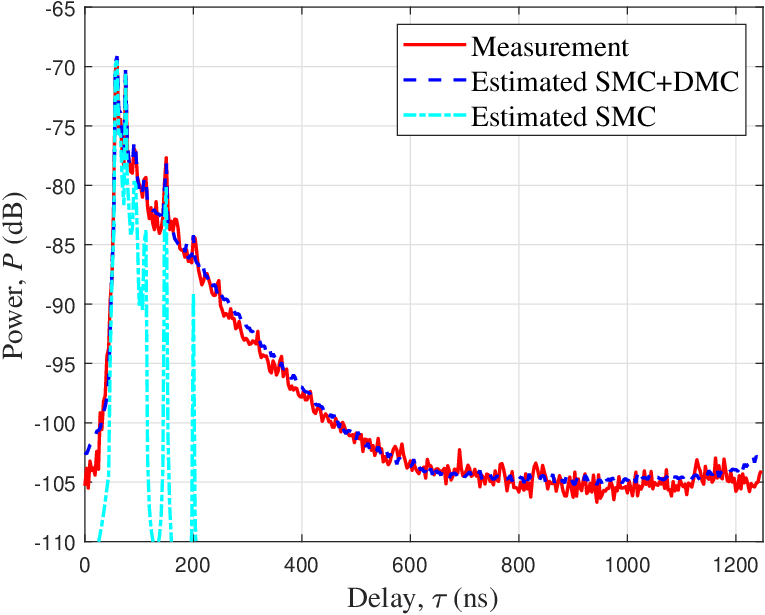}}
			\subfigure[]{\includegraphics[width = 0.7\columnwidth]{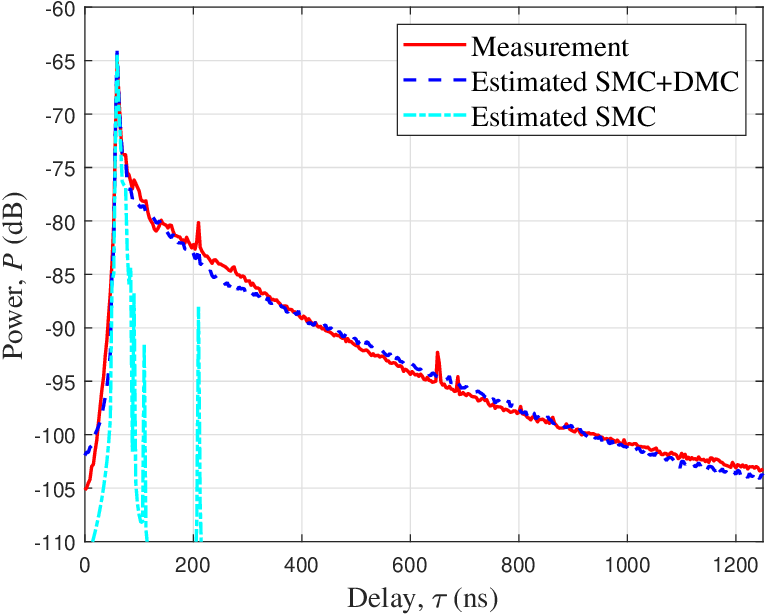}}
		\end{minipage}
		\caption{Delay PSD of measurement and estimation results at (a) position RxH35 in office scenarios and (b) position Rx40' in IIoT scenarios.}
		\label{PDP_meas_est}
	\end{figure}

	In order to compare estimation results with only SMCs and those with SMCs and DMCs to measurement results, the delay PSD averaged over all polarized antennas under the LOS condition is illustrated in Fig.~\ref{PDP_meas_est}. Several strong SMCs with powers within 20~dB of the peak power exhibit obvious peaks in delay PSDs. These reflection paths and the LOS path can be well extracted by the SAGE algorithm and account for 59.59\% and 44.91\% of the measured power in office and IIoT scenarios, respectively. The Tx--Rx distances of these two positions are comparable, while the power fraction of SMCs in the IIoT scenario is smaller than that in the office scenario, indicating the rich scattering in IIoT scenarios. Additionally, measured delay PSDs exhibit obvious long trailing and are not entirely sparse within the delay domain, which is analogous to previous studies~\cite{Zhong2020,Xu2019,Schmieder2019,Hanssens2018,Traore2018}. The trailing effect in IIoT scenarios exhibit a longer duration than that observed in office scenarios. 
	
	Considering the power proportion of SMCs and the shape of measured delay PSDs, it can be concluded that only SAGE algorithm is insufficient for extracting complete channel information. The channel is expected to comprise of both SMCs and DMCs. By combining estimation results of both SMCs and DMCs, the full average delay PSDs can closely match measurement results, which highlights the importance of the DMC modeling in the delay domain. Furthermore, the existence of DMCs may impede the detection of weaker SMCs~\cite{Zhou2023}. This is due to that the SAGE estimator relies on the successive interference cancellation and every estimated path will be subtracted from the measurement data. When powers of SMCs are lower than those of DMCs, they will not be detected. 
	  
	\subsubsection{Angular Characteristics}	
\begin{figure}[!t]
		\centering
		\begin{minipage}[!t]{\columnwidth}
			\centering
			\subfigure[]{\includegraphics[width = 0.493\columnwidth]{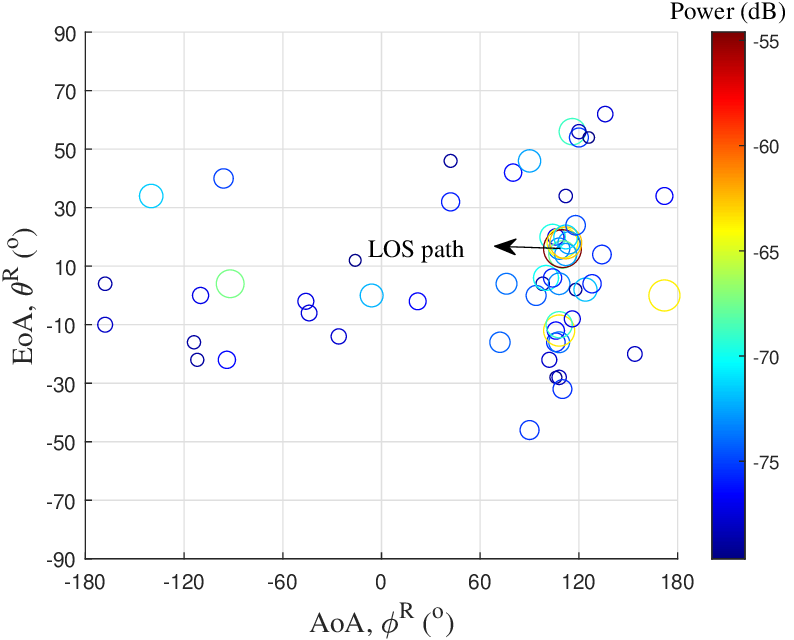}}
			\subfigure[]{\includegraphics[width = 0.493\columnwidth]{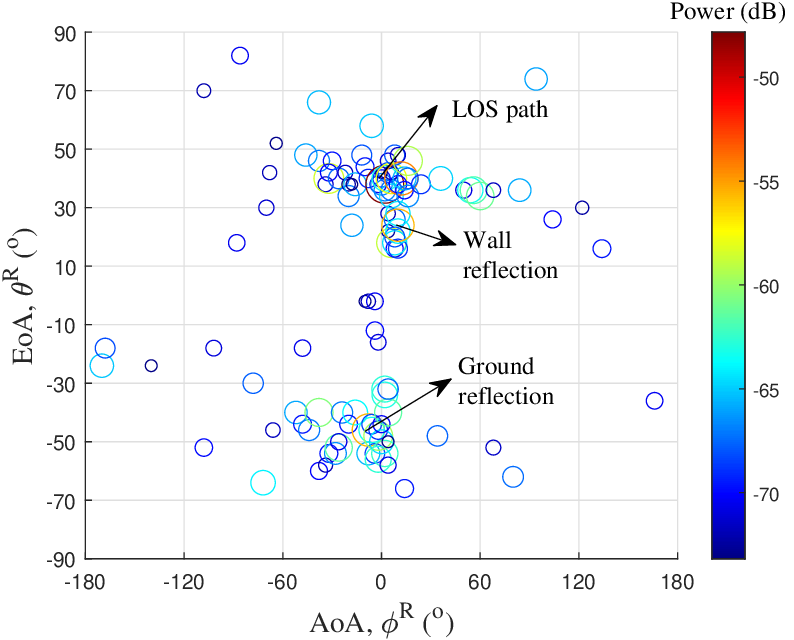}}
		\end{minipage}
		\caption{Joint angular PSD of estimation results at (a) position RxH22 in office scenarios and (b) position Rx11' in IIoT scenarios.}
		\label{angular_PSD}
	\end{figure}

	Fig.~\ref{angular_PSD} shows the joint angular PSD based on SAGE estimation results at position RxH22 in office scenarios~\cite{Zhang2023} and position Rx11' in IIoT scenarios. Only MPCs whose power is within 20~dB of the power of the LOS path are used for analysis, and powers of MPCs with same angles are combined. In IIoT scenarios, there are a number of paths with EoA less than 0$^\circ$ and the ground reflection is very obvious. For office scenarios, MPCs are around the LOS path and the ESA is smaller. 
	
	Angles of DMCs cannot be extracted as individual paths like those of SMCs. Therefore, the Bartlett beamforming method is employed to investigate angular PSDs of DMCs. The dynamic range of Bartlett beamforming results is affected by the Signal-to-Noise Ratio (SNR) and the size of the antenna array. Higher SNR and larger antenna array can increase the dynamic range. The Bartlett beamforming is calculated as
	\begin{equation}
		\begin{aligned}
			\textbf{P}_B = & \text{tr} \left( \left| \textbf{H}(\boldsymbol{f})\textbf{F}_\text{T/R}(\vartheta,\varphi)^H(\textbf{F}_\text{T/R}(\vartheta,\varphi)\textbf{F}_\text{T/R}(\vartheta,\varphi)^H)^{-1} \right.\right. \\ & \left.\left. \textbf{F}_\text{T/R}(\vartheta,\varphi)\textbf{H}(\boldsymbol{f})^H \right|  \right )
		\end{aligned}
	\end{equation}
	where $\small \textbf{H}(\boldsymbol{f}) \in \left\{ \hat{\textbf{H}}_\text{S}(\boldsymbol{f};\boldsymbol{\Theta}_\text{S}), \textbf{H}_\text{D,meas}(\boldsymbol{f}), \hat{\textbf{H}}_\text{D}(\boldsymbol{f};\boldsymbol{\Theta}_\text{D})\right\}$. Assuming a uniform spatial covariance matrix across all frequency points, as stated by~(\ref{covariance matrix of estimation}), the mean angular spectrum can be computed over $M_f$ frequency points. 
	
\begin{figure}[!t]
		\centering
		\subfigure[]{\includegraphics[width=0.75\columnwidth]{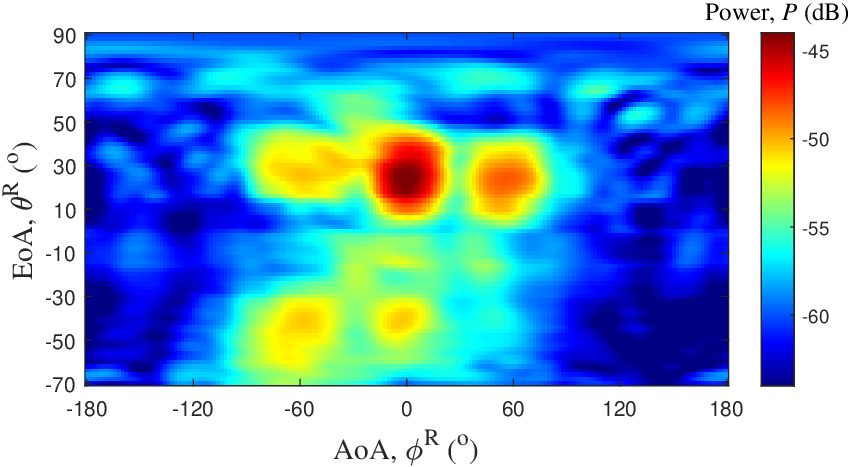}}
		\subfigure[]{\includegraphics[width=0.75\columnwidth]{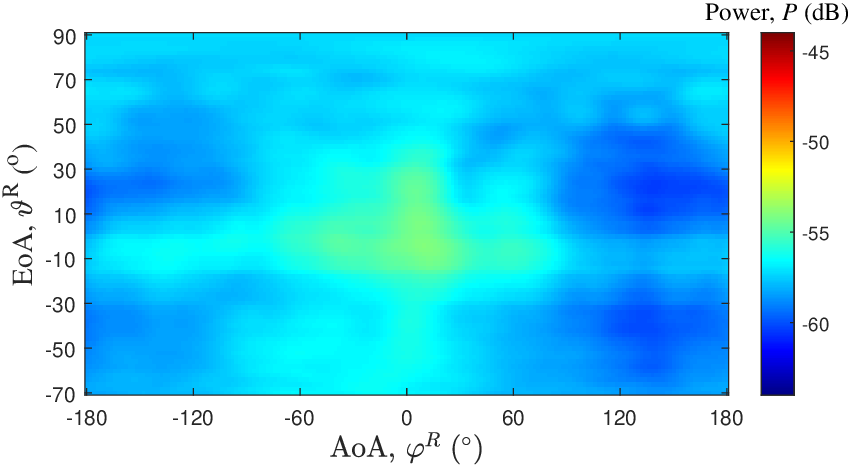}}	
		\subfigure[]{\includegraphics[width=0.75\columnwidth]{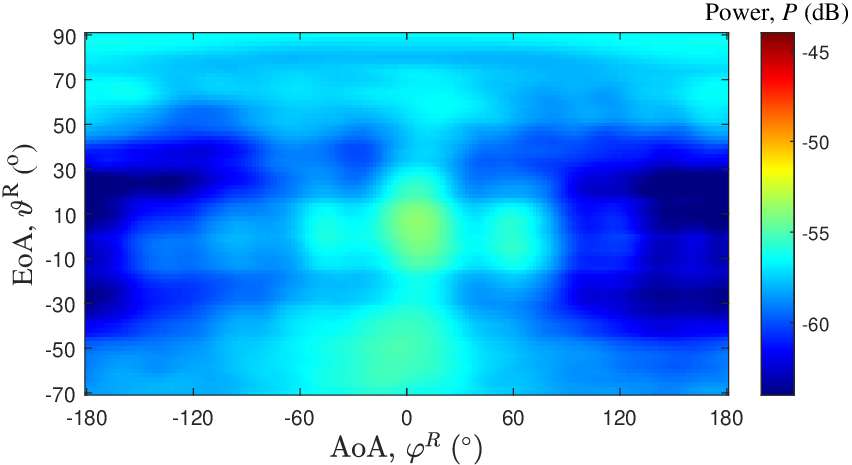}}
		\caption{Angular PSD at position Rx82' in IIoT scenarios for results of (a) estimated SMCs, (b) measured DMCs, and (c) estimated DMCs.}
		\label{PAP_meas_est}
	\end{figure}

	Fig.~\ref{PAP_meas_est} illustrates angular PSDs of estimated SMCs, measured DMCs, and estimated DMCs at position Rx82' in the IIoT scenario. Here, measured DMCs are obtained by subtracting estimated SMCs from measurement results. For an obvious comparison of their angular PSDs, the maximum dynamic range in Fig. 12(a) is set for these three subfigures. It can be found that angles of SMCs are more concentrated than those of DMCs and powers of SMCs are larger. The assumption that the angular PSD of DMCs is uniformly distributed for both departure and arrival angles is not appropriate. The energy distribution of some DMC angles is centered around SMC angles, indicating the correlation between SMCs and DMCs. In addition, the estimation results by VMF distribution exhibit a good agreement with the measurement results.
	
	ASs describe the channel dispersion in the angular domain and can be calculated by the normalized angles~\cite{Jaeckel2014}
	\begin{equation}		AS=\sqrt{\frac{\sum^L_{l=1}P_{l}\bar\phi_{l}^2}{\sum^L_{l=1}P_{l}}-\left(\frac{\sum^L_{l=1}P_{l}\bar\phi_{l}}{\sum^L_{l=1}P_{l}}\right)^2}
		\label{AS1}
	\end{equation}
	and
	\begin{equation}		
		\bar\phi_{l} = \text{angle}\left( \exp\left(j\left(\phi_{l}-\text{angle}\left(\sum_{l=1}^L\exp\left(j\phi_{l}\right)\cdot P_l\right)\right)\right)\right).
		\label{AS2}
	\end{equation}
	Here, $\phi_{l}$ can be four angles, $\bar\phi_{l}$ is the corresponding normalized angle, and $\text{angle}(\cdot)$ calculates the angle of a complex value. The logarithmic values of the four measured ASs are fitted by using the normal distribution, and their mean values $\mu_{\text{lg}(\text{AS}(^\circ))}$ and standard deviations $\sigma_{\text{lg}(\text{AS}(^\circ))}$ are presented in Table~\ref{tabAS}. The ASD is greater than ASA in both IIoT and office scenarios, because the Tx is close to the wall and numerous metal objects surround the Tx antenna, causing signal divergence in all directions. The ESD is smaller than ESA as a result of downward radiation from the Tx antenna, limiting effective angle range to half of EoA. In IIoT scenarios, ESA exceeds that of office scenarios owing to diversified scatterer heights compared to close proximity between scatterer and antenna heights in office scenarios. The measurement results exhibit significant difference with those obtained by the 3GPP channel model~\cite{3GPP}, which may be attributed to differences in the measured environments and positions.
	
	\begin{table}
		\begin{center}
			\caption{Parameters of ASs in LOS scenarios.}
			\setlength{\tabcolsep}{4pt}
			\begin{tabular}{|c|c|c|c|c|c|}			
				\hline
				\textbf{ASs}&\textbf{Parameters}& \textbf{IIoT}& \textbf{\makecell[c]{3GPP Indoor \\Factory (InF)}}& \textbf{Office}& \textbf{\makecell[c]{3GPP Indoor\\ Hotspot (InH)}}\\
				\hline
				\multirow{2}{*}{ASD ($^\circ$)}&$\mu_{\text{lg}(\text{ASD})}$& 1.78& 1.56& 1.82& 1.60\\
				& $\sigma_{\text{lg}(\text{ASD})}$& 0.1& 0.25& 0.08& 0.18 \\
				\hline
				\multirow{2}{*}{ASA ($^\circ$)}&$\mu_{\text{lg}(\text{ASA})}$& 1.46& 1.63& 1.54& 1.62\\
				& $\sigma_{\text{lg}(\text{ASA})}$& 0.14& 0.3& 0.15& 0.22\\
				\hline
				\multirow{2}{*}{ESD ($^\circ$)}&$\mu_{\text{lg}(\text{ESD})}$& 1.15& 1.35& 0.97& 1.02\\
				& $\sigma_{\text{lg}(\text{ESD})}$& 0.09& 0.35& 0.09& 0.41\\
				\hline
				\multirow{2}{*}{ESA ($^\circ$)}&$\mu_{\text{lg}(\text{ESA})}$& 1.43& 1.34& 1.03& 1.22\\
				& $\sigma_{\text{lg}(\text{ESA})}$& 0.12& 0.35& 0.15& 0.23\\
				\hline				
			\end{tabular}
			\label{tabAS}
		\end{center}
	\end{table}	
	
	\subsubsection{Power Characteristics}
\begin{figure}[!t]
		\centerline{\includegraphics[width=0.7\columnwidth]{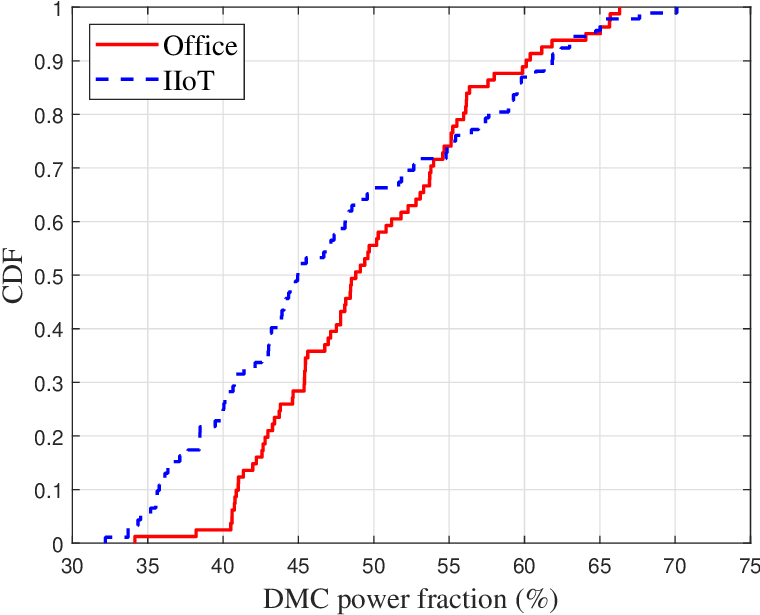}}
		\caption{The fraction of DMC powers in office and IIoT scenarios.}
		\label{DMC_power}
	\end{figure}

	Fig.~\ref{DMC_power} illustrates the power fractions of DMCs in both IIoT and office scenarios~\cite{Zhang2023} under LOS propagation conditions. The proportion of power occupied by DMCs in indoor scenarios is comparable to that of SMCs, ranging from 30\% to 70\%. It is consistently reported in~\cite{Tanghe2014,Hanssens2018,Gaillot2015} that DMCs contribute to a large ratio of the total power. Hence, the presence of DMCs cannot be disregarded in the measurement data. Furthermore, the power fraction of DMCs in office scenarios is slightly higher than that in IIoT scenarios. On one hand, the measured office environments also contain many scatterers and metal objects that can contribute to DMCs. On the other hand, the DMC power fraction is related with the Tx--Rx distance and the measured Rx positions are closer to the Tx position in IIoT scenarios compared to that in office scenarios, resulting in a lower DMC power fraction. 
	
\begin{figure}[!t]
		\centering
		\begin{minipage}[!t]{\columnwidth}
			\centering
			\subfigure[]{\includegraphics[width = 0.7\columnwidth]{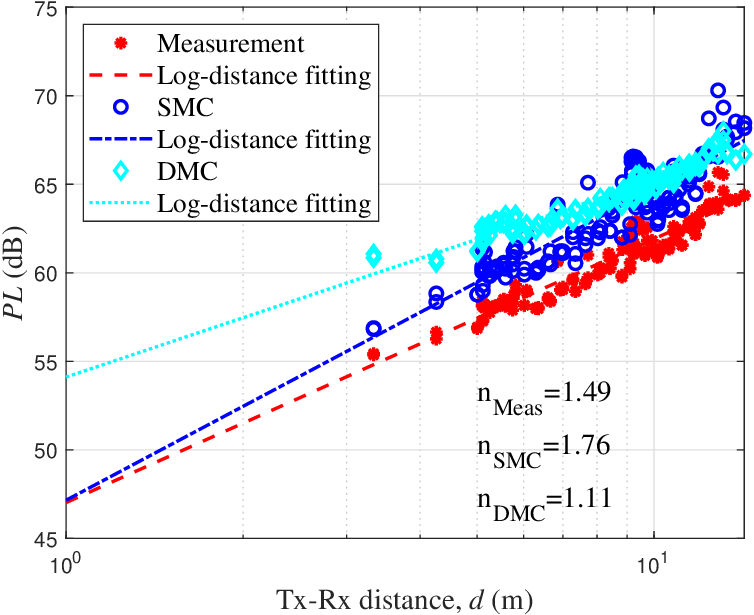}}
			\subfigure[]{\includegraphics[width = 0.7\columnwidth]{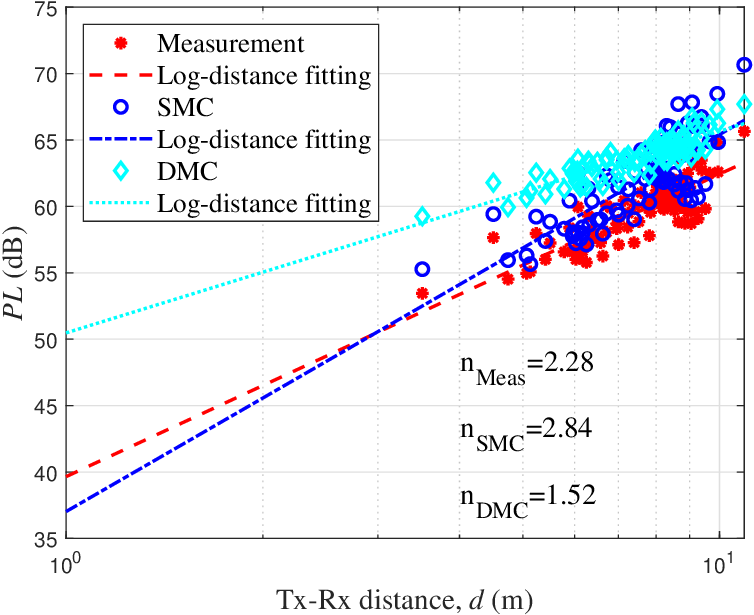}}
		\end{minipage}
		\caption{PL of measurement results, estimated SMCs, and measured DMCs in (a) office scenarios and (b) IIoT scenarios.}
		\label{PL_distance_APDP}
	\end{figure}

	Fig.~\ref{PL_distance_APDP} shows the PL of measurement results, estimated SMCs, and measured DMCs in office and IIoT scenarios. To determine MPCs from the averaged delay PSD, a threshold is set as 20 dB below the maximum power. Then, the total received power is calculated as the summation of path powers above the threshold. Different from the two-step model in~\cite{Cheng2016}, the PL is fitted only by the log-distance model. For both scenarios, the PLE of SMC results is larger while the PLE of DMC results is smaller than that of measurement results. Therefore, the power ratio of SMCs gets smaller and that of DMCs gets larger with the increase of the distance.
	\subsubsection{Polarization Characteristics}
\begin{figure}[!t]
		\centering
		\begin{minipage}[!t]{\columnwidth}
			\centering
			\subfigure[]{\includegraphics[width = 0.7\columnwidth]{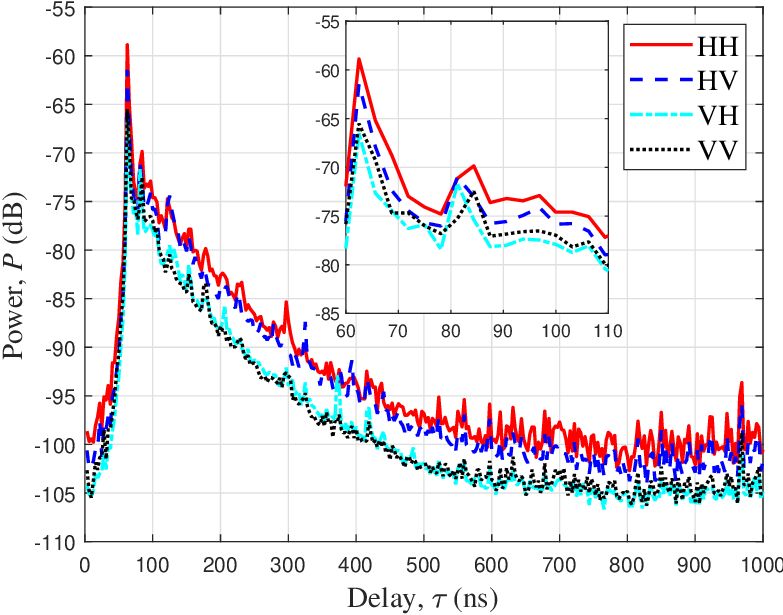}}
			\subfigure[]{\includegraphics[width = 0.7\columnwidth]{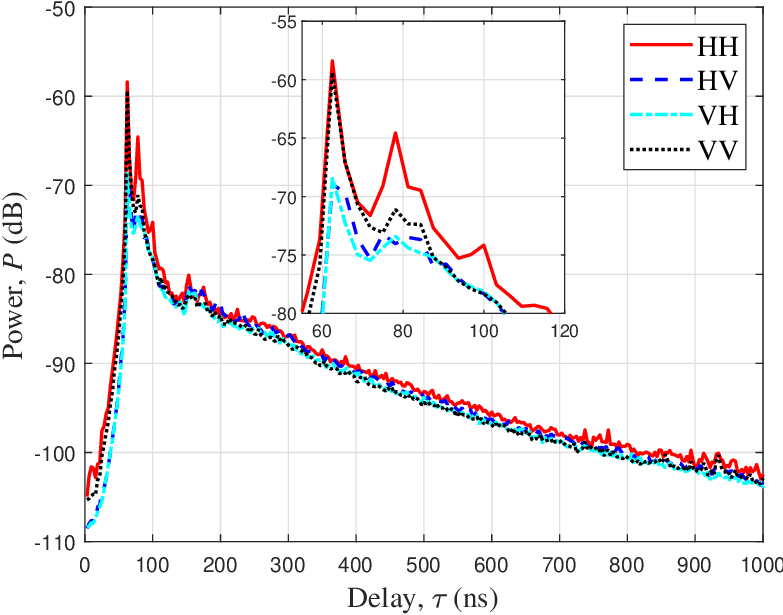}}
		\end{minipage}
		\caption{Delay PSD of measurement results for four polarization combinations at (a) position RxH11 in office scenarios and (b) position Rx20' in IIoT scenarios.}
		\label{PDP_meas_pol}
	\end{figure}

	Fig.~\ref{PDP_meas_pol} is an illustration of the delay PSD for four polarization combinations in office and IIoT scenarios. For MPCs with small delays and strong powers, the received powers of co-polarized antennas are not necessarily higher than those of cross-polarized antennas since received powers of different polarized antennas are influenced by the relative position of Tx and Rx antennas. For MPCs with larger delays and weaker powers, the horizontal-polarized Rx antenna tends to receive more powers compared to its vertical-polarized counterpart regardless of the polarization state of Tx antennas. Given that the Tx antenna is horizontally oriented, it exhibits higher radiation efficiency in the horizontal polarization direction, thereby leading to a higher received power for the horizontal-polarized Rx antenna over the vertical-polarized one. Moreover, the IIoT scenario involves pronounced ground reflections, which can enhance the power received by vertical-polarized Rx antennas. Therefore, the difference in received powers of different polarized antennas will be reduced for MPCs with larger delays and weak powers in IIoT scenarios.
	
	To investigate the statistical characteristics of the polarized received powers, XPRs for H and V polarizations as well as the Circular Polarization Ratio (CPR) are calculated as
	\begin{equation}
		\text{XPR}_\text{H} (\text{dB}) = 10\log_{10}\left(\frac{\sum_p\boldsymbol{p}_\text{HH}(\tau_p)}{\sum_p\boldsymbol{p}_\text{HV}(\tau_p)}\right)
	\end{equation} 
	\begin{equation}
		\text{XPR}_\text{V} (\text{dB}) = 10\log_{10}\left(\frac{\sum_p\boldsymbol{p}_\text{VV}(\tau_p)}{\sum_p\boldsymbol{p}_\text{VH}(\tau_p)}\right)
	\end{equation} 	
	\begin{equation}
		\text{CPR} (\text{dB}) = 10\log_{10}\left(\frac{\sum_p\boldsymbol{p}_\text{HH}(\tau_p)}{\sum_p\boldsymbol{p}_\text{VV}(\tau_p)}\right)
	\end{equation} 
	where $\boldsymbol{p}_\text{HH}(\tau_p)$, $\boldsymbol{p}_\text{HV}(\tau_p)$, $\boldsymbol{p}_\text{VH}(\tau_p)$, and $\boldsymbol{p}_\text{VV}(\tau_p)$ are delay PSDs of four polarization combinations. Table~\ref{tabXPR_MIMO} presents the normal fitting parameters for measurement results, estimated SMCs, and measured DMCs. In both scenarios, mean values of $\text{XPR}_\text{H}$ and $\text{CPR}$ are larger than 0~dB while mean values of $\text{XPR}_\text{V}$ is smaller than 0~dB. It is evident that the received powers for horizontal-polarized antennas are higher than those for vertical-polarized antennas.
	\begin{table}
		\begin{center}
			\caption{Parameters of XPR and CPR based on delay PSDs in office and IIoT scenarios.}
			\setlength{\tabcolsep}{9pt}
			\begin{tabular}{|c|c|c|c|c|c|}			
				\hline
				\multirow{2}{*}{\textbf{Components}}&\multirow{2}{*}{\textbf{XPR}}& \multicolumn{2}{c}{\textbf{Office}}& \multicolumn{2}{|c|}{\textbf{IIoT}} \\
				\cline{3-6}
				& & $\mu$&$\sigma$&$\mu$&$\sigma$\\
				\hline
				\multirow{3}{*}{Measurement}& $\text{XPR}_\text{H}$&4.23 &1.71 &1.35 &2.61 \\
				\cline{2-6}
				& $\text{XPR}_\text{V}$&-3.38 &2.41 &-0.3 &3.22\\
				\cline{2-6}
				& $\text{CPR}$&4.43 &1.14 &0.94 &1.22\\
				\hline
				\multirow{3}{*}{\makecell[c]{Estimated\\ SMC}}& $\text{XPR}_\text{H}$&4.6 &2.92 &1.4 &4.07\\
				\cline{2-6}
				& $\text{XPR}_\text{V}$&-3.77 &3.92 &-0.42 &4.53\\
				\cline{2-6}
				& $\text{CPR}$&4.93 &1.91 &0.83 &1.78\\
				\hline	
				\multirow{3}{*}{\makecell[c]{Measured\\ DMC}}& $\text{XPR}_\text{H}$&3.44 &0.39 &1.29 &0.63\\
				\cline{2-6}
				& $\text{XPR}_\text{V}$&-2.91 &0.72 &-0.68 &0.98\\
				\cline{2-6}
				& $\text{CPR}$&3.87 &0.47 &1.25 &0.46\\
				\hline					
			\end{tabular}
			\label{tabXPR_MIMO}
		\end{center}
\end{table}	

For the delay PSD, there are a lot of MPCs in each delay bin. The power at each delay bin is the fading result after multipath effects. Based on the dual-polarized SAGE algorithm, channel parameters of each path can be extracted. XPRs for H and V polarizations as well as the CPR are calculated as 
\begin{equation}		
	\text{XPR}^l_\text{H} = 10\text{log}_{10}\left(\frac{|\alpha_l^\text{HH}|^2}{|\alpha_l^\text{VH}|^2}\right)
\end{equation}	
\begin{equation}		
	\text{XPR}^l_\text{V} = 10\text{log}_{10}\left(\frac{|\alpha_l^\text{VV}|^2}{|\alpha_l^\text{HV}|^2}\right)
\end{equation}
\begin{equation}		
	\text{CPR}^l = 10\text{log}_{10}\left(\frac{|\alpha_l^\text{HH}|^2}{|\alpha_l^\text{VV}|^2}\right).
\end{equation}

In LOS scenarios, XPRs are computed separately for the LOS path and NLOS paths. In the ideal condition, the LOS path has no polarization rotation and only co-polarized powers are received. Hence, the XPR would approach infinity. However, in channel measurements, cross-polarized powers of the LOS path can be received due to various non-ideal factors. Therefore, the XPR of the LOS path describes the maximum measurable depolarization extent due to these non-ideal effects~\cite{Karttunen2015}. Table~\ref{tabXPR_SAGE} summaries the normal distribution parameters of XPR and CPR results for LOS and NLOS paths. For the LOS path, mean values of $\text{XPR}_\text{H}$ and $\text{XPR}_\text{V}$ are much larger than those for NLOS paths.The presence of numerous metal objects makes MPCs susceptible to polarization rotation. The $\text{XPR}_\text{H}$ is larger in office scenarios while $\text{XPR}_\text{V}$ is larger in IIoT scenarios, which are consistent results in Table~\ref{tabXPR_MIMO}. As mentioned before, the existence of ground reflections enhances the power received by the vertical-polarized antennas. 
\begin{table}
	\begin{center}
		\caption{Parameters of XPR and CPR based on estimated SMCs in office and IIoT scenarios.}
		\setlength{\tabcolsep}{10.5pt}
		\begin{tabular}{|c|c|c|c|c|c|}			
			\hline
			\multirow{2}{*}{\textbf{MPCs}}&\multirow{2}{*}{\textbf{XPR}}& \multicolumn{2}{c}{\textbf{Office}}& \multicolumn{2}{|c|}{\textbf{IIoT}} \\
			\cline{3-6}
			& & $\mu$&$\sigma$&$\mu$&$\sigma$\\
			\hline
			\multirow{3}{*}{LOS path}& $\text{XPR}_\text{H}$&20.58 &6.21 &17.92 &7.24 \\
			\cline{2-6}
			& $\text{XPR}_\text{V}$&13.08 &5.75 &15.74 &7.43\\
			\cline{2-6}
			& $\text{CPR}$& 1.33&3.55 & 0.45&4.32\\
			\hline
			\multirow{3}{*}{NLOS paths}& $\text{XPR}_\text{H}$&7.89 &9.85 &4.3 &9.11\\
			\cline{2-6}
			& $\text{XPR}_\text{V}$&2.25 &9.03 &2.28 &9.17\\
			\cline{2-6}
			& $\text{CPR}$&3.49 &8.62 &0.71 &8.71\\
			\hline						
		\end{tabular}
		\label{tabXPR_SAGE}
	\end{center}
\end{table}	

	\subsection{Effects of DMCs on SV and Channel Capacity}	
	\subsubsection{SV}
	The SVs of a MIMO channel matrix indicate the number of parallel independent sub-channels available and determine the capacity of each data stream. The calculations are performed in the frequency domain with the normalization factor $\gamma$ applied to all sub-channels in order to normalize the channel transfer function matrix, which is expressed by~\cite{Zhou2019}
	\begin{equation}
		\textbf{E}\left[\frac{1}{\gamma} \Vert \textbf{H}({f}) \Vert_\text{F}^2 \right] = M_\text{T}M_\text{R}
	\end{equation}
	and
	\begin{equation}
		\overline{\textbf{H}}(\boldsymbol{f}) = \frac{\textbf{H}(\boldsymbol{f})}{\sqrt\gamma}. 
	\end{equation}
	Here, $\textbf{E}\left[\cdot\right]$ is the expectation operator and $\overline{\textbf{H}}(\boldsymbol{f})$ is the normalized channel matrix.
	The SVs can be obtained through singular value decomposition at each frequency point
	\begin{equation}	
		\overline{\textbf{H}}( {f}) =  \textbf{U}\textbf{D}\textbf{V}^H.
		\label{SV}
	\end{equation}
	
\begin{figure}[!t]
		\centering
		\begin{minipage}[!t]{\columnwidth}
			\centering
			\subfigure[]{\includegraphics[width = 0.7\columnwidth]{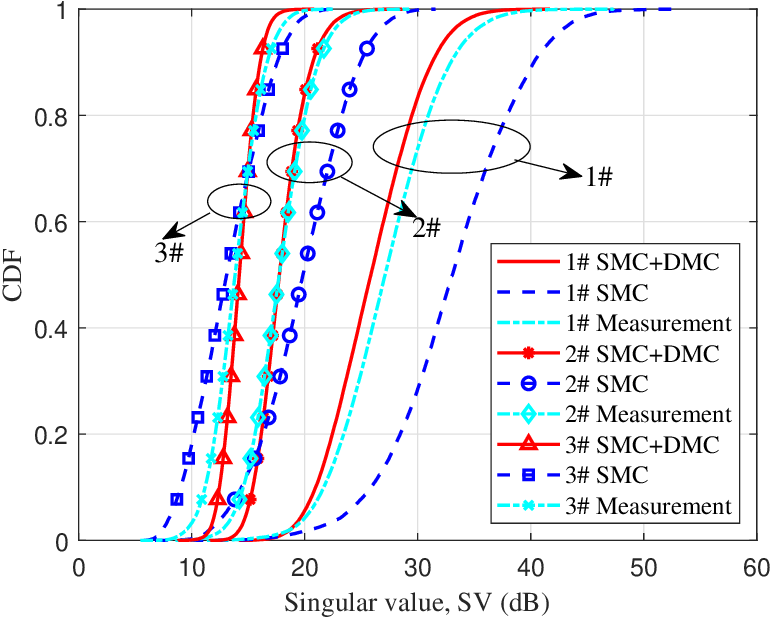}}
			\subfigure[]{\includegraphics[width = 0.7\columnwidth]{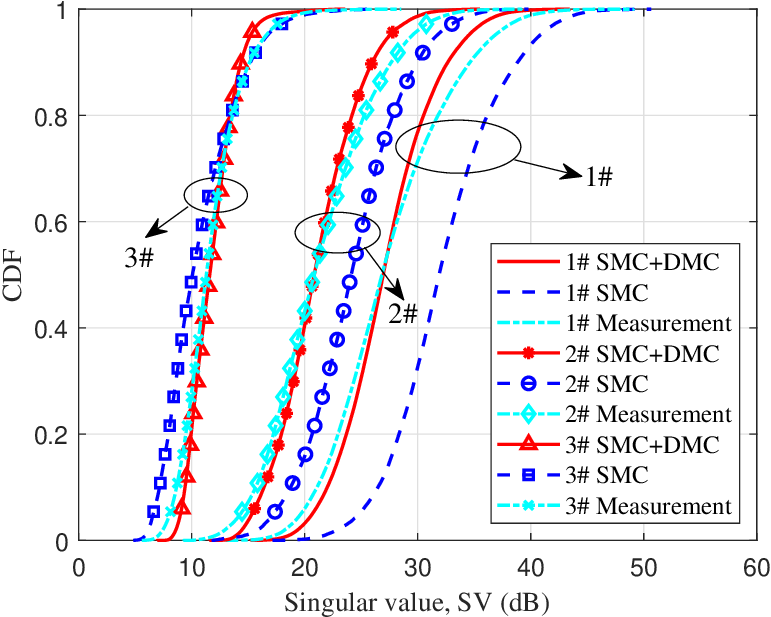}}
		\end{minipage}
		\caption{First three SVs of measurement and estimation results in (a) office scenarios and (b) IIoT scenarios.}
		\label{Three_SV}
	\end{figure}

	In~(\ref{SV}), $\textbf{U} \in \mathbb{C}^{M_\text{T} \times M_\text{T}}$ and $\textbf{V} \in \mathbb{C}^{M_\text{R} \times M_\text{R}}$ are unitary matrices, and $\textbf{D} \in \mathbb{R}^{M_\text{T} \times M_\text{R}}$ is a diagonal matrix composed of all SVs. Fig.~\ref{Three_SV} illustrates the measurement and estimation results of the first three SVs in office~\cite{Zhang2023} and IIoT scenarios, which account for 62.63\% and 68.03\% power of the channel matrix, respectively. It is evident that the disparity between the measurement results and estimations with only SMCs is greater than those with both SMCs and DMCs. This highlights once again the significance of DMCs. 

	\subsubsection{Channel Capacity}
\begin{figure}[!t]
		\centerline{\includegraphics[width=0.7\columnwidth]{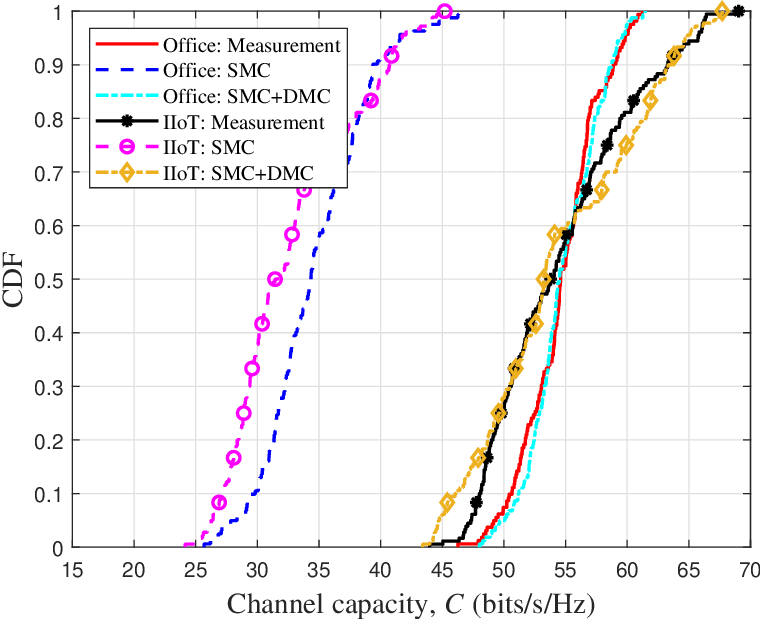}}
		\caption{Channel capacity of measurement and estimation results in office and IIoT scenarios.}
		\label{Capacity}
	\end{figure}

	The channel capacity is calculated based on the normalized channel matrix $\overline{\textbf{H}}(\boldsymbol{f})$
	\begin{equation}
		C = \textbf{E}\left[\log_2\left(\det\left|\textbf{I}_{M_\text{R}}+\frac{\rho}{M_\text{T}}\overline{\textbf{H}}(\boldsymbol{f}) \overline{\textbf{H}}(\boldsymbol{f}) ^H\right|\right)\right]
		\label{CP}
	\end{equation}
	where $\rho$ denotes the SNR, $\textbf{I}_{M_\text{R}}$ represents the identity matrix of the order $M_\text{R}$, and $\det\left|\cdot\right|$ calculates the determinant of a matrix. The channel capacity of measurement and estimation results in IIoT and office scenarios~\cite{Zhang2023} under a SNR of 5~dB is illustrated in Fig.~\ref{Capacity}. Relative errors of estimation results with only SMCs are 36.79\% and 39.82\% in office and IIoT scenarios, respectively. However, relative errors of estimation results including both SMCs and DMCs in these two scenarios are 3.68\% and 4.17\%. Neglecting the DMCs will lead to an underestimation of the channel capacity.   
	\section{Conclusions}
	\label{sec:conclusions}
	This paper has conducted both SISO and polarized MIMO channel measurements at 5.5~GHz with a bandwidth of 320~MHz in IIoT scenarios. Two typical communication scenarios, i.e., APs and industrial terminals installed on AGVs as well as those distributed across the manufacturing area, have been explored. For SISO channel measurements, various channel characteristics including the delay PSD, PL, SF, DS, ED, KF, and amplitude distribution of SSF have been investigated. Results have shown the PL in IIoT scenarios is comparatively smaller than those observed in office scenarios and free space, while the difference between PL results of Case~1 and Case~2 in IIoT scenarios is small. Besides, both DS and ED for Case~1 and Case~2 have been found to be larger in IIoT scenarios as compared to those obtained from office scenarios. Results have shown that DS and ED are larger while KF is smaller in Case~1 due to the lower antenna height compared to Case~2. In terms of MIMO channel measurements, delay, angular, power, and polarization characteristics of SMCs and DMCs have been analyzed and compared. The average delay PSD has shown multiple DMC processes in IIoT scenarios, and therefore the estimation algorithm for multiple DMC processes has been proposed based on the algorithm for a single DMC process. Angular PSDs in IIoT scenarios have been found to exhibit an obvious ground reflection. Besides, results in both scenarios have shown the existence of DMCs and the power fraction of DMCs accounts for 30\% to 70\%. In addition, the polarization status has pronounced effects on the received power and the horizontal-polarized Rx antennas tend to receive more power. Finally, the effect of DMCs on SVs and channel capacities have been explored. Results have shown that ignoring DMCs can overestimate SVs and underestimate channel capacities.

	\begin{IEEEbiography}[{\includegraphics[width=1in,height=1.25in,clip,keepaspectratio]{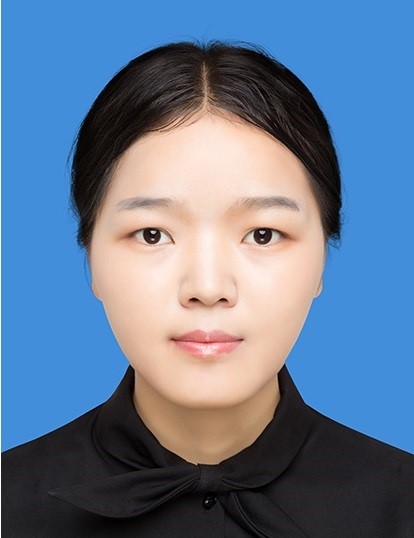}}]{Li Zhang} received the B.E. degree from Hunan University, Changsha, China, in 2019. She is currently working toward the Ph.D. degree with the National Mobile Communications Research Laboratory, Southeast University, Nanjing, China. Her research interests include indoor channel measurements, characteristics analysis, and modeling.
	\end{IEEEbiography}	
	
	\begin{IEEEbiography}[{\includegraphics[width=1in,height=1.25in,clip,keepaspectratio]{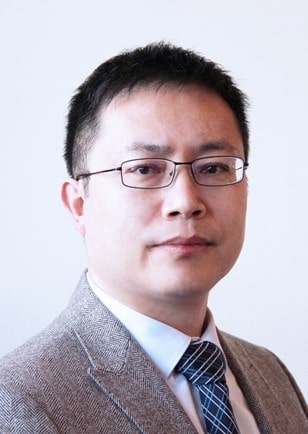}}]{Cheng-Xiang Wang} (Fellow, IEEE) received the B.Sc. and M.Eng. degrees in communication and information systems from Shandong University, China, in 1997 and 2000, respectively, and the Ph.D. degree in wireless communications from Aalborg University, Denmark, in 2004.	
		
	He was a Research Assistant with the Hamburg University of Technology, Hamburg, Germany, from 2000 to 2001, a Visiting Researcher with Siemens AG Mobile Phones, Munich, Germany, in 2004, and a Research Fellow with the University of Agder, Grimstad, Norway, from 2001 to 2005. He was with Heriot-Watt University, Edinburgh, U.K., from 2005 to 2018, where he was promoted to a professor in 2011. He has been with Southeast University, Nanjing, China, as a professor since 2018, and he is now the Executive Dean of the School of Information Science and Engineering. He is also a professor with Purple Mountain Laboratories, Nanjing, China. He has authored 4 books, 3 book chapters, and over 600 papers in refereed journals and conference proceedings, including 28 highly cited papers. He has also delivered 31 invited keynote speeches/talks and 18 tutorials in international conferences. His current research interests include wireless channel measurements and modeling, 6G wireless communication networks, and electromagnetic information theory. 
	
	Dr. Wang is a Member of the Academia Europaea (The Academy of Europe), a Member of the European Academy of Sciences and Arts (EASA), a Fellow of the Royal Society of Edinburgh (FRSE), IEEE, and IET, an IEEE Communications Society Distinguished Lecturer in 2019 and 2020, a Highly-Cited Researcher recognized by Clarivate Analytics in 2017--2020. He is currently an Executive Editorial Committee Member of the IEEE TRANSACTIONS ON WIRELESS COMMUNICATIONS. He has served as an Editor for over sixteen international journals, including the IEEE TRANSACTIONS ON WIRELESS COMMUNICATIONS, from 2007 to 2009, the IEEE TRANSACTIONS ON VEHICULAR TECHNOLOGY, from 2011 to 2017, and the IEEE TRANSACTIONS ON COMMUNICATIONS, from 2015 to 2017. He has served as a TPC Chair and General Chair for more than 30 international conferences. He received 17 Best Paper Awards from international conferences.
	\end{IEEEbiography}	
	
	\begin{IEEEbiography}[{\includegraphics[width=1in,height=1.25in,clip,keepaspectratio]{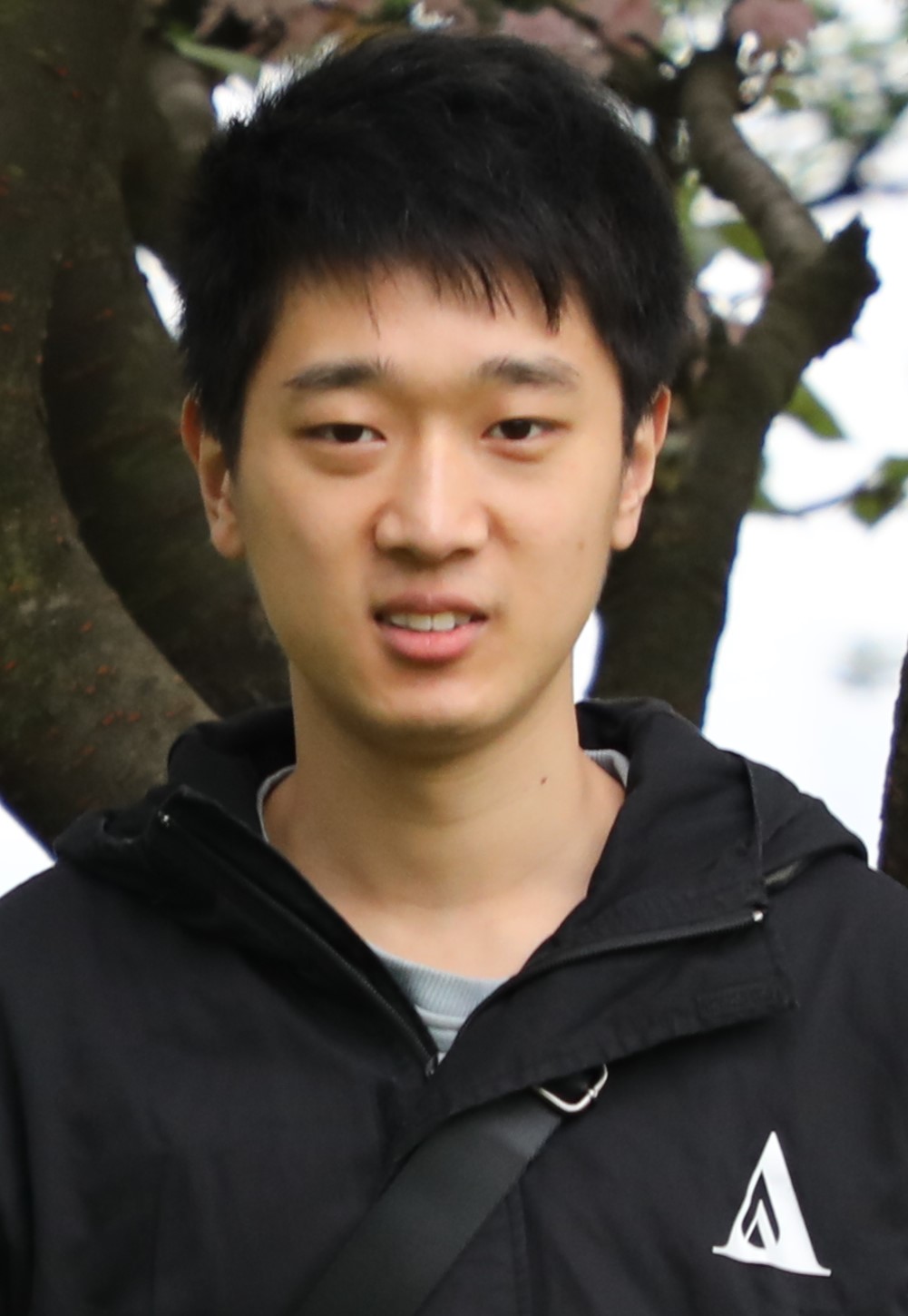}}]{Zihao Zhou} received the B.E. degree from Nanjing University of Posts and Telecommunications, Nanjing, China, in 2019. He is currently working toward the Ph.D. degree with the National Mobile Communications Research Laboratory, Southeast University, Nanjing, China. His research interests include channel parameters estimation and multi-frequency channel measurements, characteristics analysis, and modeling.
	\end{IEEEbiography}	
	
	\begin{IEEEbiography}[{\includegraphics[width=1in,height=1.25in,clip,keepaspectratio]{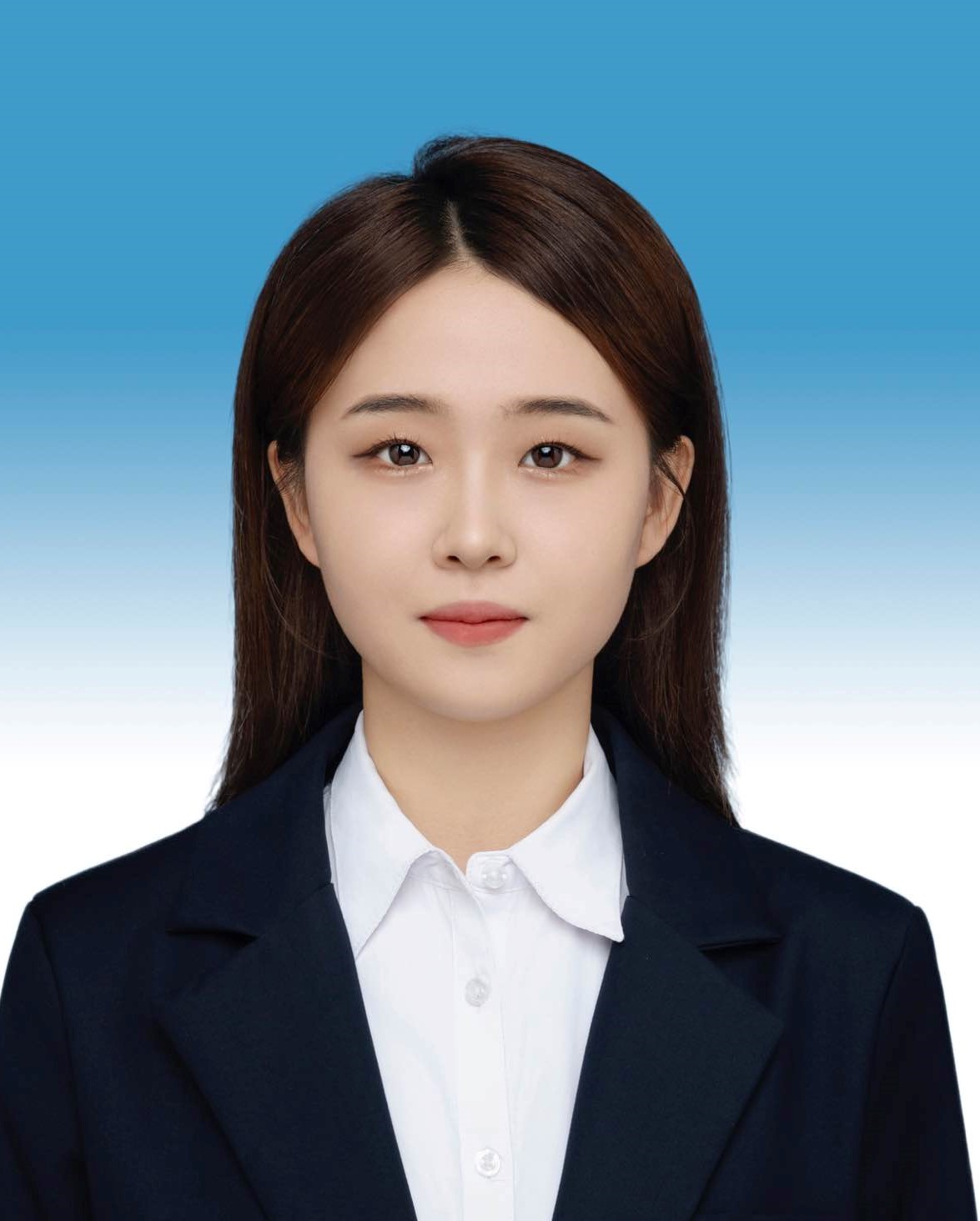}}]{Yuxiao Li} received the B.E. degree from Nanjing University of Posts and Telecommunications, Nanjing, China, in 2020, and the M.Eng degree in Information and Communication Engineering from Southeast University, Nanjing, China, in 2023. Now she works in Nanjing jiangning Power Supply Company of State Grid, Nanjing, China. Her research interests include industrial channel measurements, characteristics analysis, and modeling.
	\end{IEEEbiography}	
	
	\begin{IEEEbiography}[{\includegraphics[width=1in,height=1.25in,clip,keepaspectratio]{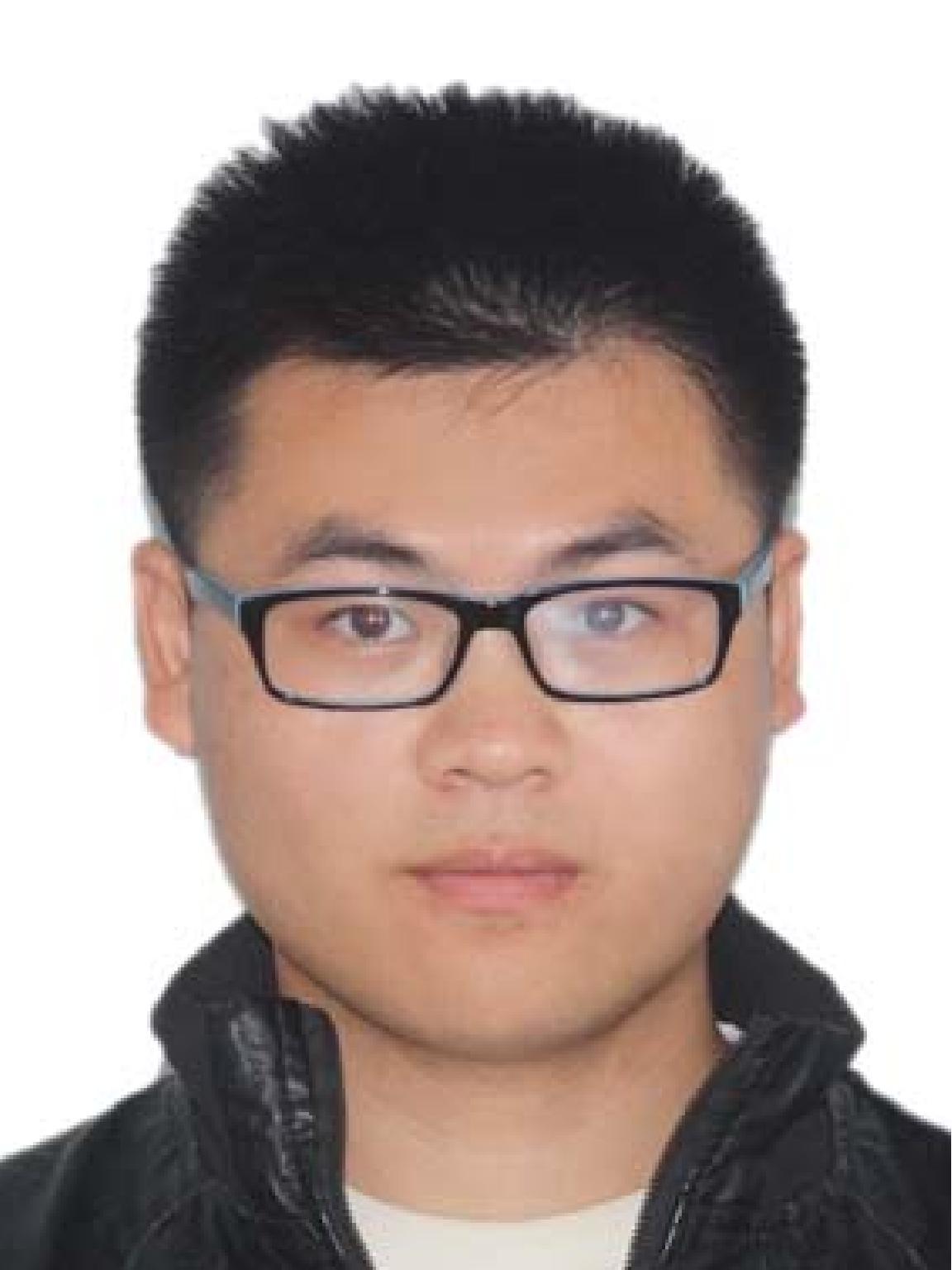}}]{Jie Huang} (Member, IEEE) received the B.E. degree in Information Engineering from Xidian University, China, in 2013, and the Ph.D. degree in Information and Communication Engineering from Shandong University, China, in 2018.

	 From Oct. 2018 to Oct. 2020, he was a Postdoctoral Research Associate in the National Mobile Communications Research Laboratory, Southeast University, China, supported by the National Postdoctoral Program for Innovative Talents. From Jan. 2019 to Feb. 2020, he was a Postdoctoral Research Associate in Durham University, UK. Since Mar. 2019, he is a part-time researcher in Purple Mountain Laboratories, China. Since Nov. 2020, he is an Associate Professor in the National Mobile Communications Research Laboratory, Southeast University. He has authored and co-authored over 100 papers in refereed journals and conference proceedings. He received the Best Paper Awards from WPMC 2016, WCSP 2020, and WCSP 2021. He has delivered 13 tutorials in international conferences. His research interests include millimeter wave, massive MIMO, reconfigurable intelligent surface channel measurements and modeling, wireless big data, electromagnetic information theory, and 6G wireless communications.
	\end{IEEEbiography}	

	\begin{IEEEbiography}[{\includegraphics[width=1in,height=1.25in,clip,keepaspectratio]{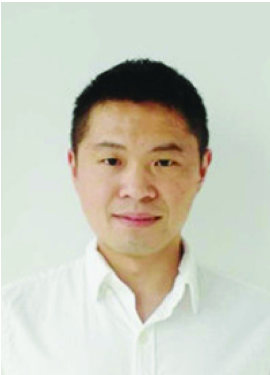}}]{Lijian Xin} (Member, IEEE) received the Ph.D. degree from Beijing University of Posts and Telecommunications (BUPT), Beijing, China, in 2021. Now he works in Purple Mountain Laboratories as an assistant research fellow, Nanjing, China. His main areas of research are over-the-air (OTA) testing of massive MIMO devices, channel measurement, and modeling.
	\end{IEEEbiography}	
	
	\begin{IEEEbiography}[{\includegraphics[width=1in,height=1.25in,clip,keepaspectratio]{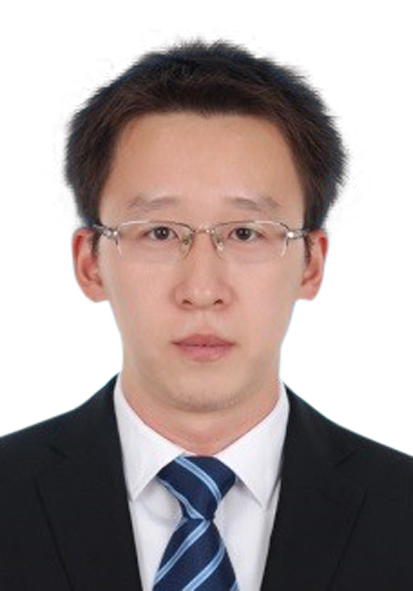}}]{Chun Pan} received his B.S. degree from Harbin Engineering University, Harbin, China, in 2010 and the Ph.D. degree from Beijing University of Posts and Telecommunications, Beijing, China, in 2015. He joined Huawei Technologies Co., Ltd., Nanjing, China, in 2015. His research interests include signal processing techniques, radio propagation, wireless sensing, ultra-high reliability transmission, and machine learning.
	\end{IEEEbiography}	
	
	\begin{IEEEbiography}[{\includegraphics[width=1in,height=1.25in,clip,keepaspectratio]{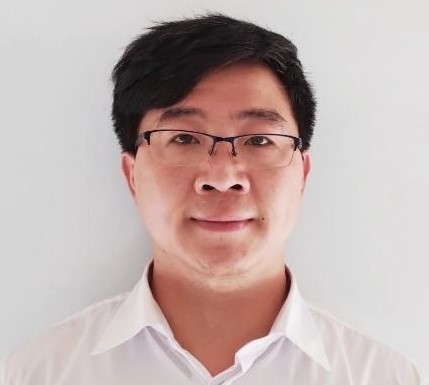}}]{Dabo Zheng} received his B.S. degree from Shenyang University of Technology, Shenyang, China, in 2013. Currently, he works as a chief engineer at Nio Inc., China. His research interests include industrial Ethernet, industrial 5G communication and secure communication protocols.
	\end{IEEEbiography}	
	
	\begin{IEEEbiography}[{\includegraphics[width=1in,height=1.25in,clip,keepaspectratio]{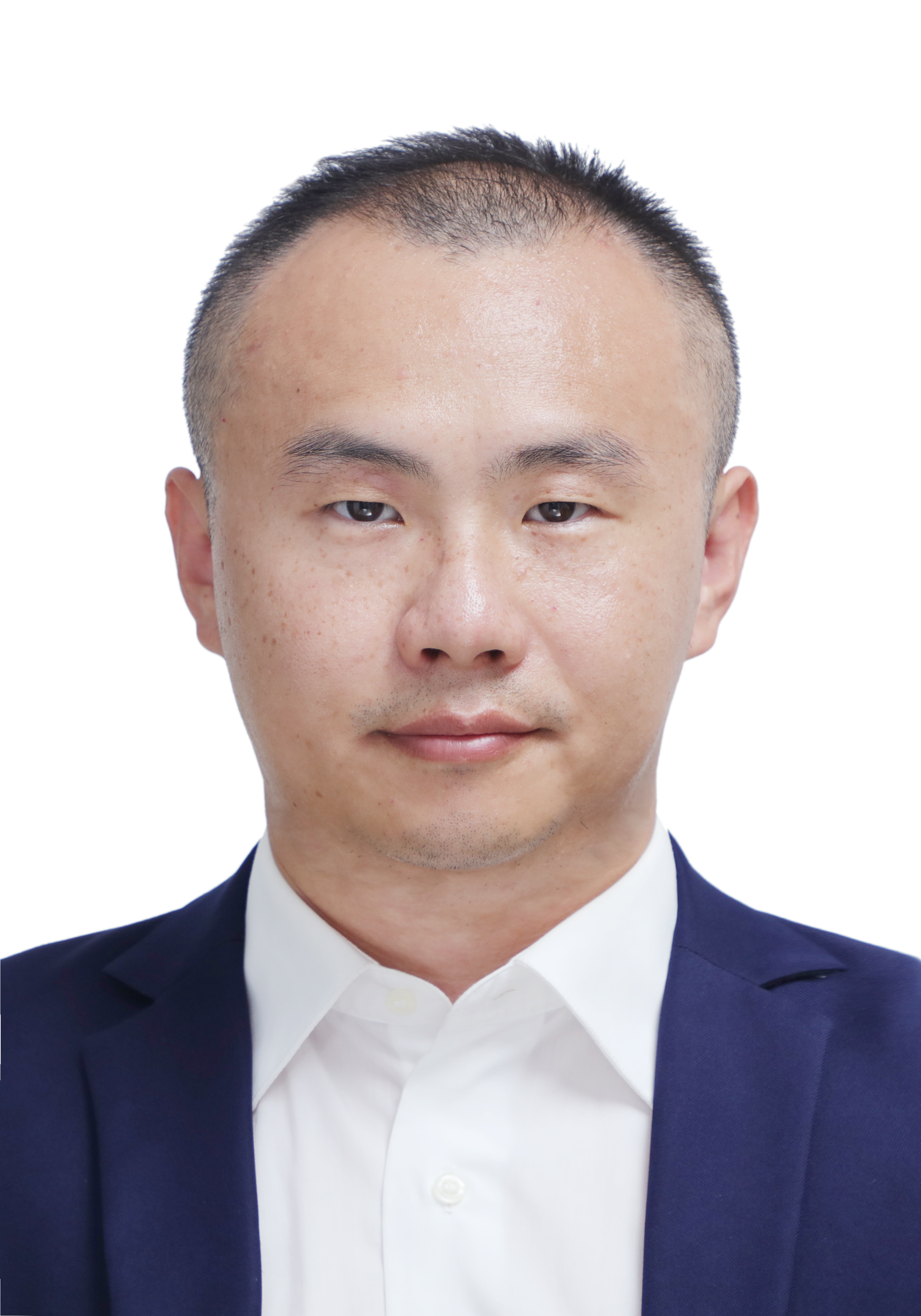}}]{Xiping Wu} (Senior Member, IEEE) received the Ph.D. degree from the University of Edinburgh, U.K., in 2015. He was a Research Associate with the University of Edinburgh, from 2015 to 2019, and a Research Fellow with Oxford University, from 2019 to 2020.  From 2020 to 2024, he was an Assistant Professor with University College Dublin, Ireland. Since 2024, he has been a professor in wireless communications with Southeast University, China. His main research interests are in the areas of 6G mobile communications, optical wireless communications (OWC), and AI-driven wireless communications. A particular focus is on developing hybrid wireless networks that integrate OWC and radio frequency, empowered by software-defined networking and AI. He has authored or coauthored over 60 papers in these research areas. He was a recipient of the Best Paper Award at IEEE ICCC in 2021 and a recipient of the Royal Irish Academy (RIA) Charlemont Grant Award in 2022.
	\end{IEEEbiography}	
		
	
\end{document}